\documentclass[english,aip,jcp,numerical,reprint,amsmath,amssymb]{revtex4-1}

\usepackage{graphicx}
\usepackage{bm}

\usepackage{color}

\newcommand{\ket}[1]{\vert#1\rangle}
\newcommand{\bra}[1]{\langle#1\vert}

\newcommand{\outt}[2]{\ket{#1}\bra{#2}}

\newcommand{\ad}{a^\dagger}

\newcommand{\sx}{\sigma_x}
\newcommand{\sz}{\sigma_z}

\newcommand{\beq}{\begin{eqnarray}}
\newcommand{\eeq}{\end{eqnarray}}

\newcommand{\tr}{\operatorname{tr}}

\begin{document}

\title{Energy transfer in structured and unstructured environments: master equations  
beyond the Born-Markov approximations}

\author{Jake Iles-Smith}
\email{Jakeilessmith@gmail.com} 
\affiliation{Controlled Quantum Dynamics Theory, Imperial College London, London SW7 2PG, United Kingdom}
\affiliation{{Photon Science Institute \& School of Physics and Astronomy, The University of Manchester, Oxford Road, Manchester M13 9PL, United Kingdom}}
\affiliation{Department of Photonics Engineering, DTU Fotonik, {\O}rsteds Plads, 2800 Kongens Lyngby, Denmark}
\author{Arend G. Dijkstra}
\affiliation{Max Planck Institute for the Structure and Dynamics of Matter, Luruper Chaussee 149, 22761, Hamburg, Germany}
\author{Neill Lambert}
\affiliation{CEMS, RIKEN, Saitama, 351-0198, Japan}
\author{Ahsan Nazir}%
\email{ahsan.nazir@manchester.ac.uk} 
\affiliation{{Photon Science Institute \& School of Physics and Astronomy, The University of Manchester, Oxford Road, Manchester M13 9PL, United Kingdom}}

\date{\today}

\begin{abstract}
We explore excitonic energy transfer dynamics in a molecular dimer system coupled to both structured and unstructured oscillator environments. By extending the reaction coordinate master equation technique developed in [J. Iles-Smith, N. Lambert, and A. Nazir, Phys. Rev. A 90, 032114 (2014)], we  
go beyond the commonly used Born-Markov approximations to incorporate system-environment correlations and the resultant non-Markovian dynamical effects. 
We obtain 
energy transfer dynamics for both underdamped and overdamped oscillator environments that are in perfect agreement with the numerical hierarchical equations of motion over a wide range of parameters. 
Furthermore, we show that the Zusman equations, which may be obtained in a semiclassical limit of the reaction coordinate model, are often incapable of describing the correct dynamical behaviour. This demonstrates the necessity of properly accounting for quantum correlations generated between the system and its environment 
when the Born-Markov approximations no longer hold.
Finally, we apply the reaction coordinate formalism to the case of a structured environment comprising of both underdamped (i.e.~sharply peaked) and overdamped (broad) components simultaneously. 
We find that though an enhancement of the dimer energy transfer rate can be obtained 
when compared to an unstructured environment, its magnitude is rather sensitive to both 
the dimer-peak resonance conditions and the relative strengths of the underdamped and overdamped contributions.
\end{abstract}
\maketitle

\section{Introduction}
\label{sec:sec1}

Since the first observations of 
coherent signatures 
in photosynthetic systems~\cite{engel2007evidence,Lee07,collini10,panitchayangkoon10}, determining whether such effects play 
a functional role in promoting efficient and robust excitonic energy transfer (EET) across pigment-protein complexes has been a driving force for the field of quantum biology~\cite{Fleming:2004aa,revishizaki,Lambert:2013aa}.
Theoretical work on the subject quickly identified that purely coherent energy transport is insufficient to explain the high efficiencies and rates observed \cite{mohseni:174106,PlenioHuelga,ishizakipnas}.
For resonant systems, this is due to the inherent reversibility of coherent dynamics, while biased systems become localised in the site basis when the inter-site energy difference is greater than the tunnelling energy, thus reducing exciton transport. 
These difficulties may be circumvented when noise processes induced by an external environment are also present, providing mechanisms for rapid and efficient EET~\cite{mohseni:174106,PlenioHuelga,ishizakipnas,caruso:105106,1367-2630-12-6-065002,1367-2630-12-10-105012,1367-2630-11-3-033003,Mintert_review}. 

However, accurately accounting for the effects of the external environment in photosynthetic systems is a daunting theoretical prospect. Strong coupling between the system and its 
environment leads to the accumulation of significant system-environment correlations that may be present even within the steady-state~\cite{noncan,cklee12,PhysRevA.90.032114,cerrillo14}. In addition to the (often low frequency) 
continuum, the environmental spectral densities of pigment-protein complexes are generally structured; that is, there are specific underdamped vibrational modes of the environment that couple strongly to the excitonic degrees of freedom.
There is now increasing evidence to suggest that these underdamped modes are an important contributing factor to the long-lived 
coherences observed in photosynthetic systems~\cite{PhysRevA.90.012510,ishizaki2,Levi,Kolli12,christensson12,Chin13,Plenio13,turner12,richards12,tiwari13,chenu13,richards14,romero14,fuller14,ryu14,novelli2015vibronic,PhysRevLett.105.050404,womick11,womick3}, as well as 
links between the quantum mechanical nature of these vibrational modes and enhanced transfer rates~\cite{womick09,womick11,Chin13,OReilly:2014aa}.

A multitude of powerful   
computational methods have been developed to deal with the difficulties faced in modelling strongly dissipative quantum systems. Examples include the hierarchical equations of motion (HEOM)~\citep{Tanimura2,tan1,Tan2,Tan3,doi:10.1021/jz502701u,HEOMfit}, density matrix renormalisation group (and related) techniques~\cite{PhysRevLett.105.050404,Chin13,schroder2015simulating,rosenbach15}, and those based on the path integral formalism~\cite{makri95a,Makri1992435,Thorwart2009234,nalbach:194111}. All can converge to numerically exact results in specific circumstances. 
In contrast, despite their attraction in terms of simplicity, intuitive physical insight and efficiency, standard (e.g.~Redfield) master equations are often invalid in regimes relevant to molecular complexes due to their limitation to weak system-environment couplings ~\cite{ishizaki:234110,breuer2007theory}. Though procedures such as the polaron transformation can be used to broaden the range of validity of Redfield master equations~\cite{PhysRevLett.103.146404,PhysRevB.83.165101,varET,felix13,Kolli11,jang08,jang09,jang11,zimanyi12}, they may again be restricted; for example to situations in which the high-frequency environmental response dominates~\cite{varET,PhysRevB.84.081305,cklee12b,chang13}. 

\begin{figure}[t]
\includegraphics[width=0.49\textwidth]{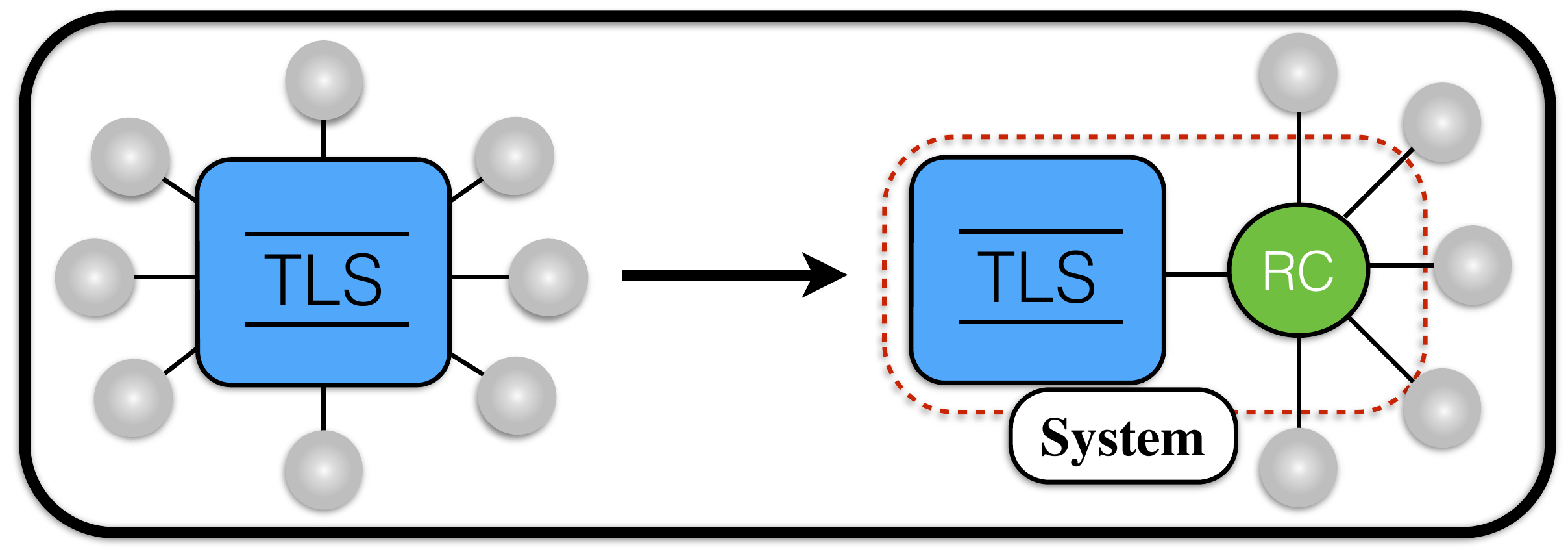}
\caption{Schematic of the reaction coordinate mapping, showing a two-level-system (TLS) coupled directly to an oscillator environment (left), which is mapped to a TLS coupled to the reaction coordinate (RC) plus a residual bath (right).}
\label{fig:schem}
\end{figure}

Recently, a master equation approach based on the reaction coordinate (RC) model (see Fig.~\ref{fig:schem}) was introduced to describe the dynamical behaviour of a system coupled to an environment with strong low-frequency components, leading to long environmental correlation times~\cite{PhysRevA.90.032114}.
Here, a collective coordinate of the bath 
~\cite{woods,PhysRevA.50.3650,garg:4491,thoss:2991, cao97, hughes:124108, hartmann:11159,martinazzo:011101,PhysRevLett.115.130401,pollard1} is incorporated into an enlarged effective system Hamiltonian. 
This allows 
for the derivation of a second-order master equation for the dynamics of 
the reduced system and RC, 
accurately describing the system dynamics even in the presence of strong system-environment correlations and extended environmental memory. Apart from its conceptual simplicity, the reaction coordinate master equation (RCME) is attractive due to the additional insight that it provides beyond the system, 
into both the environmental dynamics and the generation of 
system-environment correlations~\cite{PhysRevA.90.032114}. 

In this work, 
we shall employ 
the RCME to investigate EET in a molecular dimer system beyond weak system-environment coupling. 
We shall show that over the broad regimes considered, the RC model captures all important system-environment correlations 
for EET in the presence of both underdamped and overdamped environments, agreeing perfectly with numerically exact data generated using the HEOM~\cite{tan1,Tan2,Tan3,doi:10.1021/jz502701u}.
Furthermore, we demonstrate that the RC model significantly outperforms the closely related Zusman equations~\cite{thoss:2991,hartmann:11159,Yijang}, a set of drift-diffusion equations often used to describe tunnelling processes in molecular systems, which we derive from the RCME in a semiclassical limit. 
We also examine the role that a structured environment may play in the dimer energy transfer dynamics. 
This is 
achieved in a consistent and non-perturbative manner by incorporating an underdamped mode into the system Hamiltonian using the RC formalism, 
while the broad background environment is described using a second overdamped RC.
We show that the presence of structure in the spectral density can enhance the EET rate in specific regimes, in particular when the characteristic frequency of the underdamped environment coincides with the excitonic resonance of the molecular dimer. 
However, there are also large regions of parameter space where no enhancement is to be expected.

The paper is organised as follows. In Section~\ref{sec:sec2} we define the molecular dimer Hamiltonian and outline the RC mapping. In Section~\ref{sec:sec3} we formulate the RCME, from which we also derive the well-established Zusman equations~\cite{thoss:2991,hartmann:11159,Yijang}. 
In Section~\ref{sec:sec4} we 
explore the dynamics of the model for both overdamped and underdamped environmental spectral densities 
using the RCME and Zusman equations, 
which we benchmark against the HEOM technique. 
In Section~\ref{sec:sec5} we extend our discussion 
to a structured environment, with a particular focus on the effect of environmental structure on the rate of energy transfer between the dimer sites. We summarise in Section~\ref{sec:sec6} and present further details of the Zusman equations in the Appendix.

\section{System Hamiltonian and the reaction coordinate mapping}
\label{sec:sec2}

We consider energy transport in a molecular dimer system, each site of which has an excited state $\ket{X_j}$ $(j=1,2)$ with an associated energy $\varepsilon_j$, and ground state $\ket{0_j}$. The two sites are coupled to each other via a transition dipole interaction, with strength $\Delta$, and 
to separate 
harmonic environments, leading to a Hamiltonian of the form (where we set $\hbar=1$)
\begin{align}\label{eq:rawham}
H&=\sum\limits_j \varepsilon_j \ket{X_j}\bra{X_j}+\frac{\Delta}{2}\left(\ket{X_10_2}\bra{0_1X_2}+\ket{0_1X_2}\bra{X_10_2}\right)\nonumber\\
&+\sum\limits_j\ket{X_j}\bra{X_j}\sum\limits_k f_{j,k}\left(c_{j,k}^\dagger+c_{j,k}\right)+\sum\limits_{j,k}\omega_{j,k}c_{j,k}^\dagger c_{j,k}.
\end{align}
Here, $c_{j,k}^\dagger$ and $c_{j,k}$ are, respectively, the creation and annihilation operators for the  $k^{th}$ mode of the environment at site $j$, and $f_{j,k}$ is the corresponding coupling strength. 
To simplify the analysis, 
we assume that the couplings between each site and its environment are identical, then rotate the coordinate system 
such that the two sites couple directly to a single bath within the single excitation subspace  
spanned by the basis $\left\{\ket{1}=\ket{X_1 0_2},\ket{2}=\ket{0_1X_2}\right\}$. 
Restricting ourselves to the single exciton subspace allows us treat the dimer 
as an effective two level system (TLS), and we can then write the Hamiltonian in spin-boson form
\begin{align}\label{eq:SB}
{H}_{\rm SB}=\frac{\epsilon}{2} \sigma_{z}+\frac{\Delta}{2}\sigma_x&+\sigma_z\sum\limits_k f_k\left(\tilde{c}_k^\dagger+\tilde{c}_k\right)+\sum\limits_k\nu_k \tilde{c}_k^\dagger \tilde{c}_k,
\end{align}
where $\epsilon=\epsilon_1-\epsilon_2$, $\tilde{c}_k=\frac{1}{2}\left(c_{1,k}-c_{2,k}\right)$, and 
$f_k=f_{1,k}/\sqrt{2}$. We have also introduced the Pauli operators, $\sigma_z=\vert 1\rangle\langle 1\vert-\ket{2}\bra{2}$ and $\sigma_x=\ket{2}\bra{1}+\ket{1}\bra{2}$.
We can characterise the system-bath interaction by introducing the spectral density, $J_{\rm SB}(\omega)=\sum_k \vert f_k\vert^2\delta(\omega-\omega_k)$, which is a measure of coupling strength 
weighted by the environmental density of states. 
 
Given that our Hamiltonian is now in spin-boson form, the derivation of the RCME 
proceeds as in Ref.~\onlinecite{PhysRevA.90.032114}, which we shall summarise here for completeness. 
To move beyond the weak-coupling limit appropriate to Born-Markov (e.g~Redfield) master equations we apply a normal mode transformation to Eq.~(\ref{eq:SB}), incorporating a collective environmental degree of freedom into a new effective system Hamiltonian. 
Following the method outlined by Garg {et al.}~\cite{garg:4491}, 
we define a collective mode of the environment, known as the RC, 
which couples directly to the TLS. The RC is in turn coupled to a residual harmonic environment, as can be seen schematically in Fig.~\ref{fig:schem}. 
This leads to a Hamiltonian of the form
 \begin{align}\label{eq:RC}
H_{\rm RC}&=H_{\rm S}+H_{\rm I}+H_{\rm B}+H_{\rm C},
\end{align}
with
\begin{align}
H_{\rm S}&=\frac{\epsilon}{2} \sigma_{z}+\frac{\Delta}{2}\sigma_x+\lambda\sigma_z\left(a^\dagger+a\right)+\Omega a^\dagger a,\nonumber\\
H_{\rm I}&=\left(a^\dagger+a\right)\sum\limits_k g_k\left(b_k^\dagger+b_k\right),\nonumber\\
H_{\rm B}&=\sum\limits_k\omega_k b^\dagger_kb_k,\nonumber\\
H_{\rm C}&=\left(a^\dagger+a\right)^2\sum\limits_k\frac{g_k^2}{\omega_k},\nonumber
\end{align}
 where the collective coordinate is defined such that 
 \begin{equation}
 \lambda\left(\hat{a}^\dagger+\hat{a}\right)=\sum\limits_kf_k\left(\tilde{c}_k^\dagger+\tilde{c}_k\right).
 \end{equation}
In Eq.~(\ref{eq:RC}) we have added a term, $H_{\rm C}$, quadratic in the position operator of the RC, which removes the renormalisation of the mode potential due to friction~\cite{hartmann:11159}. 
We have also defined new creation and annihilation operators, $\hat{b}_k^\dagger$ and $\hat{b}_k$, respectively, for the residual bath. This couples directly to the RC and is characterised by a new spectral density, $J_{\rm RC}(\omega)=\sum_k\vert g_k\vert^2\delta(\omega-\omega_k)$. 

\begin{figure}[t]
\center
	\includegraphics[width=0.24\textwidth]{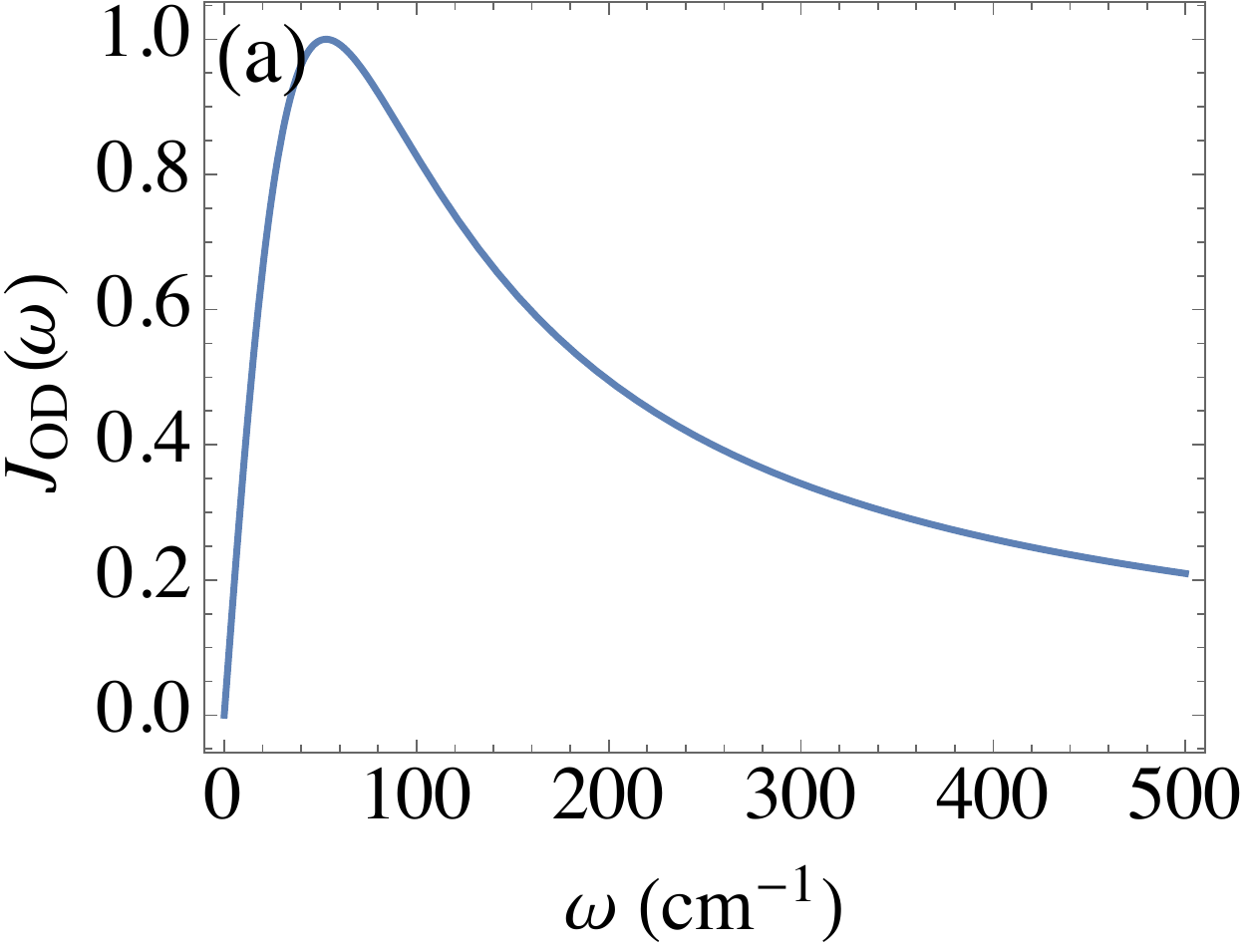}
	\includegraphics[width=0.235\textwidth]{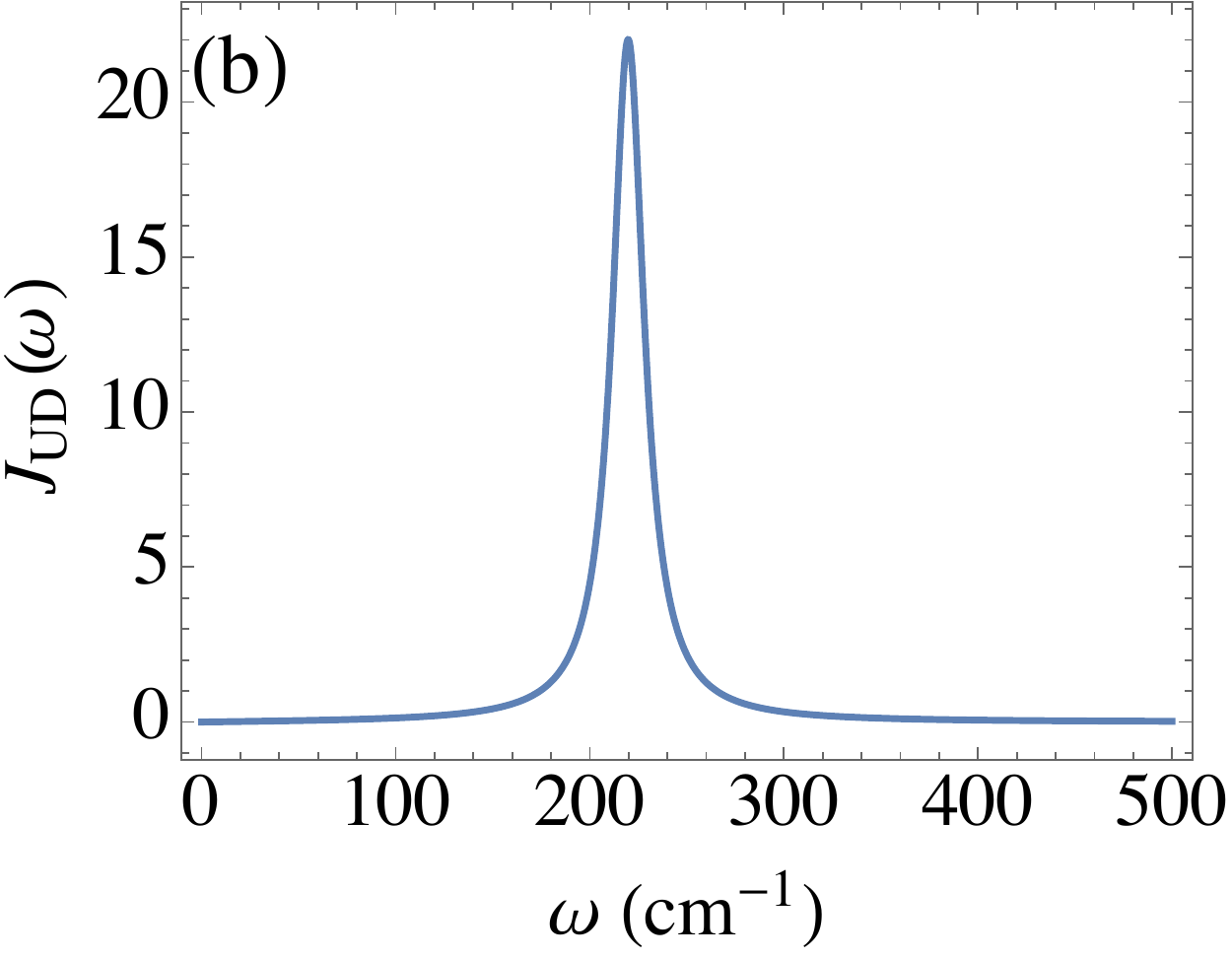}
	\caption{Example spectral densities considered in this work: (a) Overdamped spectral density with $\omega_c =53$~cm$^{-1}$ and $\pi\alpha =2$~cm$^{-1}$; (b) Underdamped spectral density with $\pi\alpha =2$~cm$^{-1}$, $\Gamma=20$~cm$^{-1}$, and $\omega_0 = 220$~cm$^{-1}$.}
	\label{fig:sds}
\end{figure}

To describe the action of the residual bath on the RC we need to relate this spectral density to the original spin-boson spectral density $J_{\rm SB}(\omega)$. To do so, we replace the TLS with a classical coordinate $q$ subject to a potential $V(q)$~\cite{garg:4491,hughes:124108}. 
 By considering the Fourier transformed equations of motion for the coordinate, both before and after the mapping, we gain expressions of the form~\cite{PhysRevA.90.032114}
\begin{equation}
\tilde{K}(z)\tilde{q}(z)=-\tilde{V}^\prime(q),
\end{equation}
where tildes refer to Fourier transforms and prime denotes the derivative with respect to $q$. 
For example, after the RC mapping, the Fourier space operator may be written as
\begin{equation}
\tilde{K}(z)=-z^2+\frac{2\lambda}{\Omega}\frac{\mathcal{L}(z)}{\Omega^2+\mathcal{L}(z)},
\end{equation}
with $\mathcal{L}(z)=-z^2-4\Omega z^2\int^\infty_0 \frac{J_{\rm RC}(\omega)}{\omega(\omega^2-z^2)}d\omega$. 
Finally, the spin-boson spectral density may be related to $J_{\rm RC}(\omega)$ 
using the Leggett prescription~\cite{PhysRevB.30.1208}:
\begin{equation}\label{eq:map}
J_{\rm SB}(\omega)=\frac{1}{\pi}\lim_{\epsilon\rightarrow0+}{\rm Im}\left(\tilde{K}(\omega-i\epsilon)\right).
\end{equation}
In the following we shall study two spectral densities relevant to EET 
systems, the underdamped (UD) and overdamped (OD) Brownian oscillator forms
\begin{align}\label{jsbud}	
J_{\rm SB}^{\rm UD}(\omega) =&\frac{ \alpha_{\rm UD} \Gamma \omega_0^2 \omega}{(\omega_0^2-\omega^2)^2 + \Gamma^2 \omega^2},
\end{align}
and
\begin{align}\label{jsbod}	
J_{\rm SB}^{\rm OD}(\omega)=&\alpha_{\rm OD}\omega_c\frac{\omega}{\omega^2+\omega_c^2}.
\end{align}
By choosing the RC spectral density to have the form $J_{\rm RC}(\omega)= \gamma\omega\exp\left(-\omega/\Lambda\right)$, and using Eq.~(\ref{eq:map}) in the limit that $\Lambda\rightarrow\infty$, we find the relation
\begin{equation}
J_{\rm SB}(\omega)=\frac{4\gamma\omega\Omega^2\lambda^2}{(\Omega^2-\omega^2)^2+(2\pi\gamma\Omega\omega)^2}.
\end{equation}
Thus the mapping described above is exact for the underdamped spectral density when $\Omega=\omega_0$, $\lambda=\sqrt{\pi \alpha_{\rm UD}\omega_0 / 2}$, and $\gamma=\Gamma/2\pi\omega_0$. We can also recover the overdamped spectral density by choosing $\gamma$ such that $\omega_c\ll\Omega$, where the RC coupling strength and frequency satisfy
\begin{alignat}{2}
 \omega_c &= \frac{\Omega}{2\pi\gamma}  &\qquad\text{and}\qquad  \alpha_{\rm OD} &= \frac{2\lambda^2}{\pi\Omega}.
\end{alignat}
Fig.~\ref{fig:sds} gives illustrative examples of the spectral densities defined in Eqs.~(\ref{jsbud}) and~(\ref{jsbod}), demonstrating that in comparison to the broader overdamped limit, the underdamped case displays a sharp peak centred about the characteristic frequency $\omega_0$. A combination of the two will be used in Section~\ref{sec:sec5} below to represent a structured spectral density, arising from coupling of the dimer to both its background continuum environment and a specific lossy mode of vibration.

\section{Dynamical evolution in the RC model}
\label{sec:sec3}
By mapping the spin-boson Hamiltonian of Eq.~(\ref{eq:SB}) to the RC form given in Eq.~(\ref{eq:RC}), we are now in a position 
to proceed with the dynamical description of our dimer system. 
We shall consider two related approaches. In the first we derive the full RCME (a quantum master equation) for the {reduced dynamics of both the dimer and the RC}, 
while in the second we apply further approximations  
in order to derive a set of partial differential equations (PDEs), known as the Zusman equations.

\subsection{Reaction coordinate master equation}

We consider a second-order 
master equation for the reduced state of the RC and TLS, $\rho(t)$. This accounts for the TLS-RC coupling exactly, while treating 
interactions with the residual environment perturbatively to second order. 
This treatment will be valid when the coupling between the mapped system and the residual environment is weak or when the residual environment correlation time is short.
From Eq.~(\ref{eq:RC}), moving into the interaction picture with respect to $H_{\rm S}+H_{\rm B}$, we may write the Born-Markov master equation as~\cite{breuer2007theory}
\begin{equation}\label{eq:me}
\begin{split}
\frac{\partial\rho_{\rm I}(t)}{\partial t}=&-i\left[H_{\rm C}(t),\rho(0)\right]\\
&-\int\limits_0^\infty d\tau\ {\rm tr}_{\rm B} \left[H_{\rm I}(t),\left[H_{\rm I}(t-\tau),\rho_{\rm I}(t)\otimes\rho_{\rm B}\right]\right],
\end{split}
\end{equation}
where $\rho_{\rm B}=e^{-\beta H_{\rm B}}/{\rm tr}_{\rm B} \{e^{-\beta H_{\rm B}}\}$ is the reduced state of the residual bath, which is assumed to remain in thermal equilibrium at temperature $T=1/\beta$ (for $k_B=1$). This assumption is justified 
when the coupling between the residual bath and the composite system is small or the residual bath correlation time is very short. By following the derivation outlined in Ref.~\onlinecite{PhysRevA.90.032114}, we may write the Schr\"odinger picture master equation for the combined TLS and RC as
\begin{equation}\label{eq:mastertext}
\begin{split}
\dot{\rho}(t)&=-i\left[H_{\rm S},\rho(t)\right]\\
&-\gamma\int\limits_0^\infty d\tau\int\limits_0^\infty  d\omega\ \omega\cos\omega\tau\coth\frac{\beta\omega}{2} \left[\hat{A},\left[\hat{A}(-\tau),\rho_{\rm S}(t)\right]\right]\\
&-\gamma\int\limits_0^\infty d\tau\int\limits_0^\infty  d\omega \cos\omega\tau\left[\hat{A},\left\{ \left[\hat{A}(-\tau),H_{\rm S}\right],\rho(t)\right\}\right],
\end{split}
\end{equation}
where we have defined $\hat{A}=\hat{a}^\dagger+\hat{a}$.

The complexity of the system Hamiltonian makes gaining an analytic expression for the interaction picture operators difficult. 
However, by simply truncating the space of the collective coordinate up to $n$ basis states, i.e. limiting ourselves to $n$ excitations in the RC, we can numerically diagonalise $H_{\rm S}$. 
This approach leads to the set of basis states $\ket{\varphi_j}$ which satisfy the relation $H_{\rm S}\ket{\varphi_j}=\varphi_j\ket{\varphi_j}$, allowing us to write the interaction picture operators as
\begin{equation}
\hat{A}(t)=\sum\limits_{j,k=1}^{2n}A_{jk}e^{i\omega_{jk}t}\ket{\varphi_j}\bra{\varphi_k},
\end{equation}
where $A_{jk}=\bra{\varphi_j}{\hat{a}^\dagger+\hat{a}}\ket{\varphi_k}$ and $\omega_{jk}=\varphi_j-\varphi_k$. We can now evaluate the time and frequency integrals in Eq.~(\ref{eq:mastertext}) to give
\begin{equation}\label{eq:RCME}
\begin{split}
\frac{\partial\rho(t)}{\partial t}=&-i\left[H_{\rm S},\rho(t)\right]-\left[\hat{A},\left[\hat{\chi},\rho(t)\right]\right]+\left[\hat{A},\left\{\hat\Xi,\rho(t)\right\}\right],
\end{split}
\end{equation}
where we have defined the rate operators 
\begin{align}
\hat{\Xi}&=\frac{\pi}{2}\sum\limits_{j,k=1}^{2n}\gamma\omega_{jk}A_{jk}\ket{\varphi_j}\bra{\varphi_k},\label{eqn:Xi}\\
\hat{\chi}&=\frac{\pi}{2}\sum\limits_{j,k=1}^{2n}\gamma\omega_{jk}\coth\frac{\beta\omega_{jk}}{2}A_{jk}\ket{\varphi_j}\bra{\varphi_k},\label{eqn:chi}
\end{align}
and assumed the 
imaginary parts 
(i.e.~Lamb shifts) 
to be negligible. Eq.~(\ref{eq:RCME}) thus captures the interaction between the TLS and RC non-perturbatively, while the residual bath is treated in a purely Markovian fashion. 

\subsection{Zusman Equations}

From the  RCME given in Eq.~(\ref{eq:mastertext}), we can derive a set of drift-diffusion PDEs by way of further approximations. 
Specifically, the 
interaction picture operators in Eq.~(\ref{eq:mastertext}) are expanded using the Caldeira-Leggett 
approach~\cite{breuer2007theory,Caldeira1983587,thoss:2991,garg:4491}, in which the 
system evolution is assumed to be much slower 
than that of the environment, giving 
\begin{align}
\hat{A}(t)=e^{-iH_{\rm S} t}\hat{A}e^{iH_{\rm S} t}\approx\hat{A}+it\left[H_{\rm S},\hat{A}\right].
\end{align}
By inserting this approximate form 
into Eq.~(\ref{eq:mastertext}) we can evaluate the frequency and time integrals.
We then move to a phase-space representation for the master equation by way of the Wigner transformation,
leading to a generalised Fokker-Plank equation in Klein-Kramers form~\cite{Risken:1984aa}
\begin{align}\label{eq:wigner}
\frac{\partial\hat{W}}{\partial t}+\mathcal{H}\hat{W}+&(i\mathcal{Z}-\Omega^2x)\frac{\partial\hat{W}}{\partial p}+p\frac{\partial\hat{W}}{\partial x} \nonumber\\&= \pi\gamma\Omega\frac{\partial}{\partial t}\left(p\hat{W}+\frac{1}{\beta}\frac{\partial\hat{W}}{\partial p}\right),
\end{align}
where $\hat{W}=\sum_{ij}W_{ij}(x,p,t)\ket{i}\bra{j}$, with $i,j=1,2$. 
The Wigner function is defined as 
\begin{align}
W_{ij}(x,p,t)=\frac{1}{2\pi}\int\limits_{-\infty}^\infty dx^\prime e^{-i p x^\prime}\left\langle i, x+\frac{x^\prime}{2}\right\vert \rho\left\vert x-\frac{x^\prime}{2},j\right\rangle,
\end{align}
where $x$ and $p$ are the phase-space coordinates of the RC. For brevity, we have defined the superoperators in Eq.~(\ref{eq:wigner}) as
 \begin{align}
  \mathcal{H}\hat{W} &=i\left[(\epsilon+\sqrt{2\Omega}\lambda x)\sigma_z+\Delta\sigma_x,\hat{W}\right],\\
 \mathcal{Z}\frac{\partial\hat{W}}{\partial p}&=\frac{i\sqrt{2\Omega}\lambda}{2}\left\{\sigma_z,\frac{\partial\hat{W}}{\partial p}\right\}.
 \end{align}
 
In its current form Eq.~(\ref{eq:wigner}) remains challenging to solve. We may simplify it, however, 
by eliminating the momentum coordinate in the differential equation.
We do this by assuming that the RC momentum remains in thermal equilibrium at all times, which is valid in the high friction limit. This enables us to expand the Wigner function in terms of Hermite polynomials, resulting in a hierarchy of equations (a detailed account of this derivation can be found in the Appendix). 
By taking terms that are first order in the inverse friction ($\eta^{-1}$, where $\eta=\pi\gamma\Omega$),  
we acquire a set of drift-diffusion equations, commonly referred to as the Zusman equations:

\begin{widetext}
\begin{align}\label{eq:ZUS}
\frac{\partial\mu_{11}(t,x)}{\partial t}&=\frac{1}{2\pi\gamma\Omega}\frac{\partial}{\partial x}\left(\frac{1}{\beta}\frac{\partial\mu_{11}(t,x)}{\partial x}+(\Omega^2 x+\sqrt{2\Omega}\lambda)\mu_{11}(t,x)\right)+i\Delta(\mu_{12}(t,x)-\mu_{21}(t,x)),\\
\frac{\partial\mu_{22}(t,x)}{\partial t}&=\frac{1}{2\pi\gamma\Omega}\frac{\partial}{\partial x}\left(\frac{1}{\beta}\frac{\partial\mu_{22}(t,x)}{\partial x}+(\Omega^2 x+\sqrt{2\Omega}\lambda)\mu_{22}(t,x)\right)-i\Delta(\mu_{12}(t,x)-\mu_{21}(t,x)),\\
\frac{\partial\mu_{12}(t,x)}{\partial t}&=\frac{1}{2\pi\gamma\Omega}\frac{\partial}{\partial x}\left(\frac{1}{\beta}\frac{\partial\mu_{12}(t,x)}{\partial x}+\Omega^2 x\mu_{12}(t,x)\right)
+2i(\frac{\epsilon}{2} +\sqrt{2\Omega}\lambda x)\mu_{12}(t,x)+i\frac{\Delta}{2}(\mu_{11}(t,x)-\mu_{22}(t,x)),\\
\frac{\partial\mu_{21}(t,x)}{\partial t}&=\frac{1}{2\pi\gamma\Omega}\frac{\partial}{\partial x}\left(\frac{1}{\beta}\frac{\partial\mu_{21}(t,x)}{\partial x}+\Omega^2 x\mu_{21}(t,x)\right)
+2i(\frac{\epsilon}{2} +\sqrt{2\Omega}\lambda x)\mu_{21}(t,x)-i\frac{\Delta}{2}(\mu_{11}(t,x)-\mu_{22}(t,x)),
\end{align}
\end{widetext}
where $\mu_{ij}(t,x)$ 
describes the time evolution of both the TLS and the RC with respect to the phase-space variable 
$x$.
We can then extract the time evolution of the TLS population [$\rho_{11}(t)$] and coherence [$\rho_{12}(t)$] using
\begin{equation*}
\rho_{11}(t)=\int\limits_{-\infty}^\infty \mu_{11}(t,x)dx\hspace{0.45cm}\text{and}\hspace{0.45cm}\rho_{12}(t)=\int\limits_{-\infty}^\infty \mu_{12}(t,x)dx.
\end{equation*}
The Zusman equations describe a mode in the high friction limit and are 
based on approximations that amount to a semiclassical treatment of the RC, in which quantum correlations between the RC and dimer are neglected.
One may extend their validity by considering higher order terms. However, 
the equations quickly become unwieldy and computationally impractical in the low friction limit. Thus, we shall restrict ourselves to the first-order equations here.

\section{System dynamics}\label{sec:sec4}

To explore the system dynamics using the RCME [Eq.~(\ref{eq:RCME})], we assume that the dimer and RC are initially uncorrelated at time $t=0$, with the RC in a thermal state and an excitation localised at dimer site $1$ (unless otherwise stated). That is, $\rho(0)=\mathcal{Z}_0^{-1}\ket{1}\bra{1}\otimes\exp\left(-\beta\Omega a^\dagger a\right)$, where $\mathcal{Z}_0=\tr\left\{\exp\left(-\beta\Omega a^\dagger a\right)\right\}$.  
For the Zusman equations, this gives the boundary condition
\begin{equation}
\mu_{11}(0,x)=2\sqrt{\frac{\tanh\left(\frac{\beta\Omega}{2}\right)}{\pi}}e^{-\Omega\tanh\left(\frac{\beta\Omega}{2}\right)x^2},
\end{equation}
while $\mu_{22}(0,x)=\mu_{21}(0,x)=\mu_{12}(0,x)=0$.
 We shall compare the dynamical behaviour predicted by the RCME and Zusman equations, solved numerically~\cite{Johansson20121760,Johansson20131234}, for the overdamped and underdamped spectral densities in turn. 
 This will be benchmarked against the HEOM. As the HEOM are derived from the original (i.e.~unmapped) Hamiltonian of Eq.~(\ref{eq:SB}), the appropriate initial state is an excitation localised at site $1$ with the environment in a multimode thermal state at the same temperature as defined in the RC case.

\subsection{Overdamped spectral density}

\begin{figure}[t]\center
 \includegraphics[width=0.2385\textwidth]{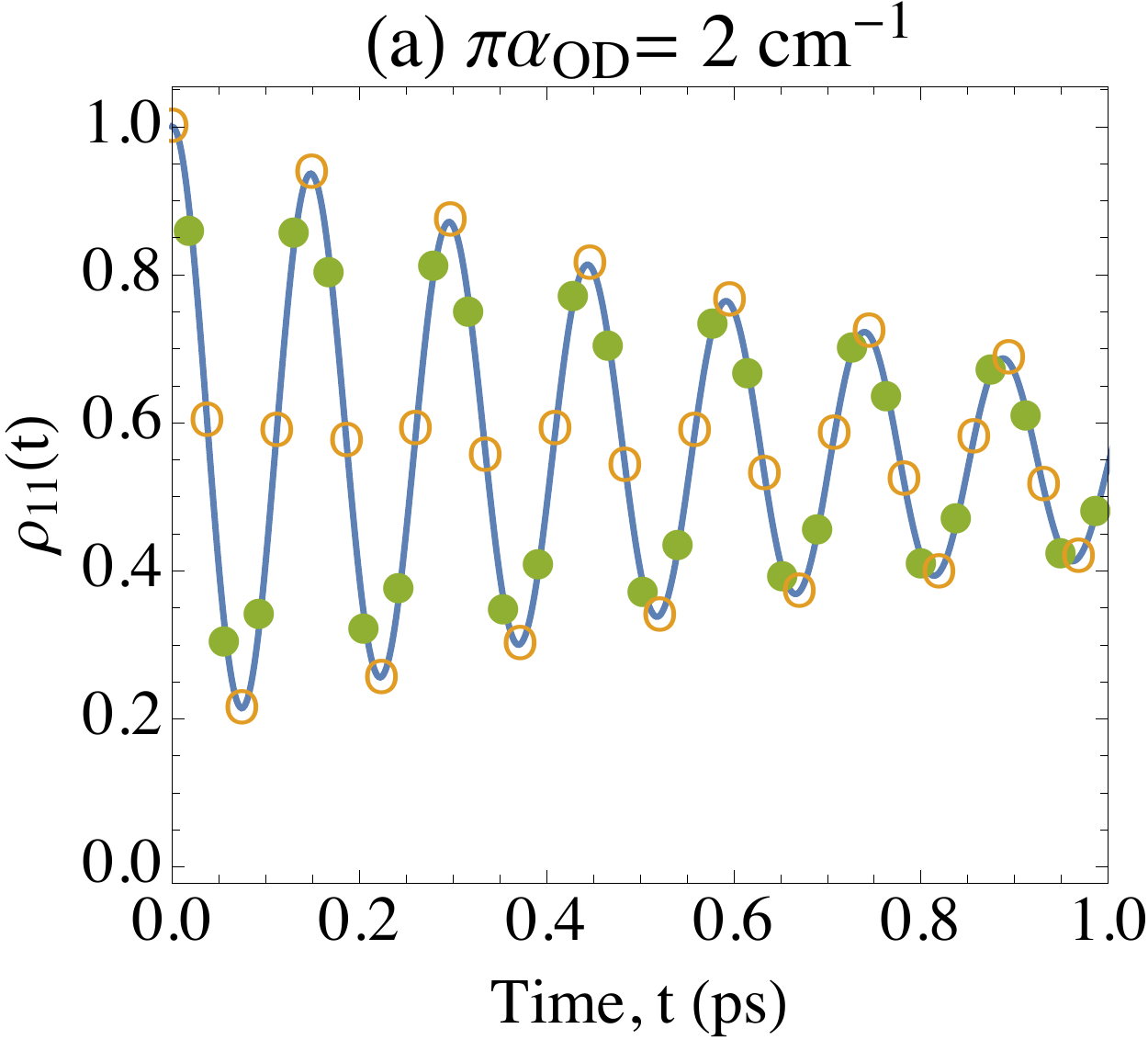}
 \includegraphics[width=0.2385\textwidth]{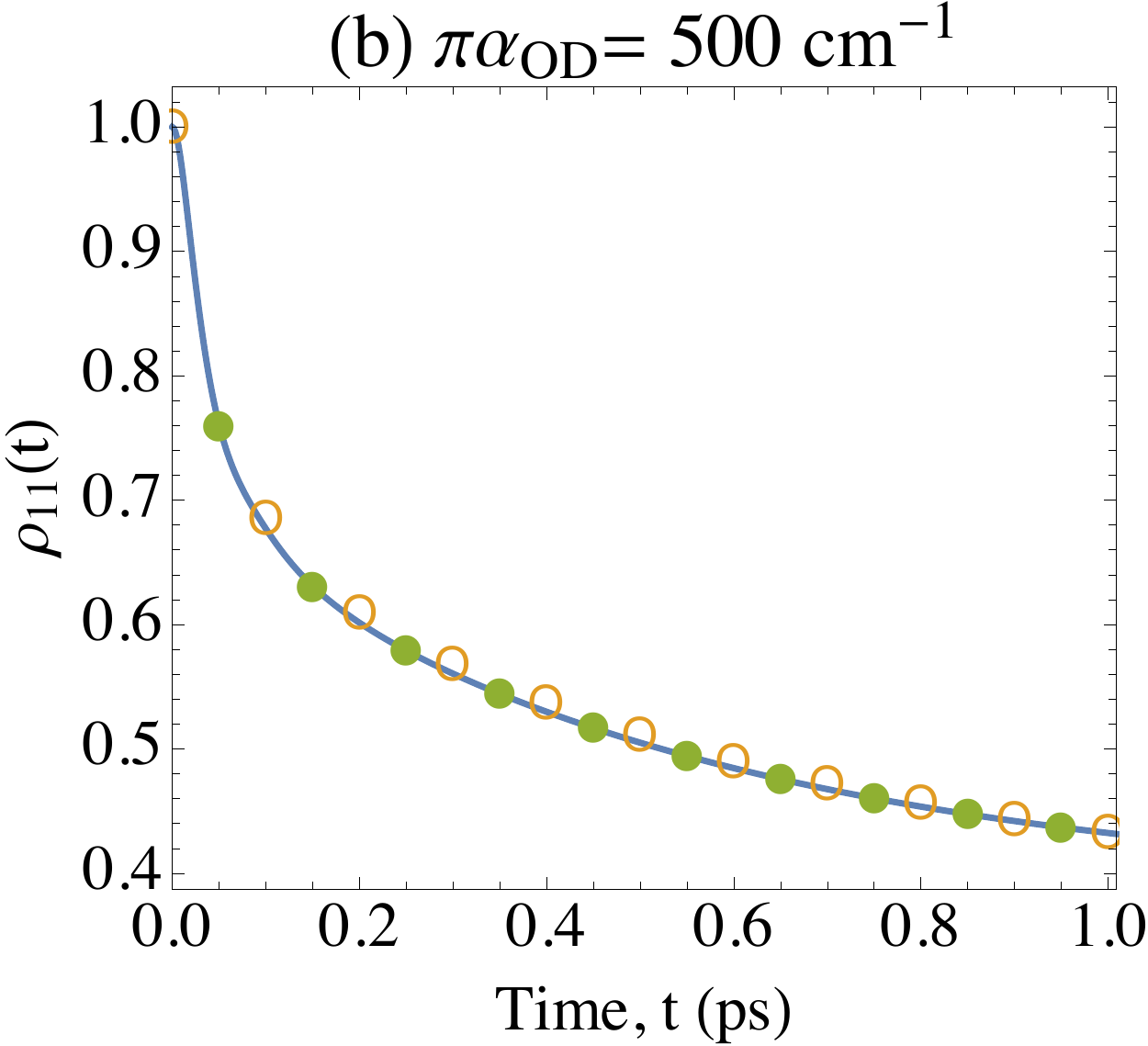}\\ \vspace{0.1cm}
  \includegraphics[width=0.2385\textwidth]{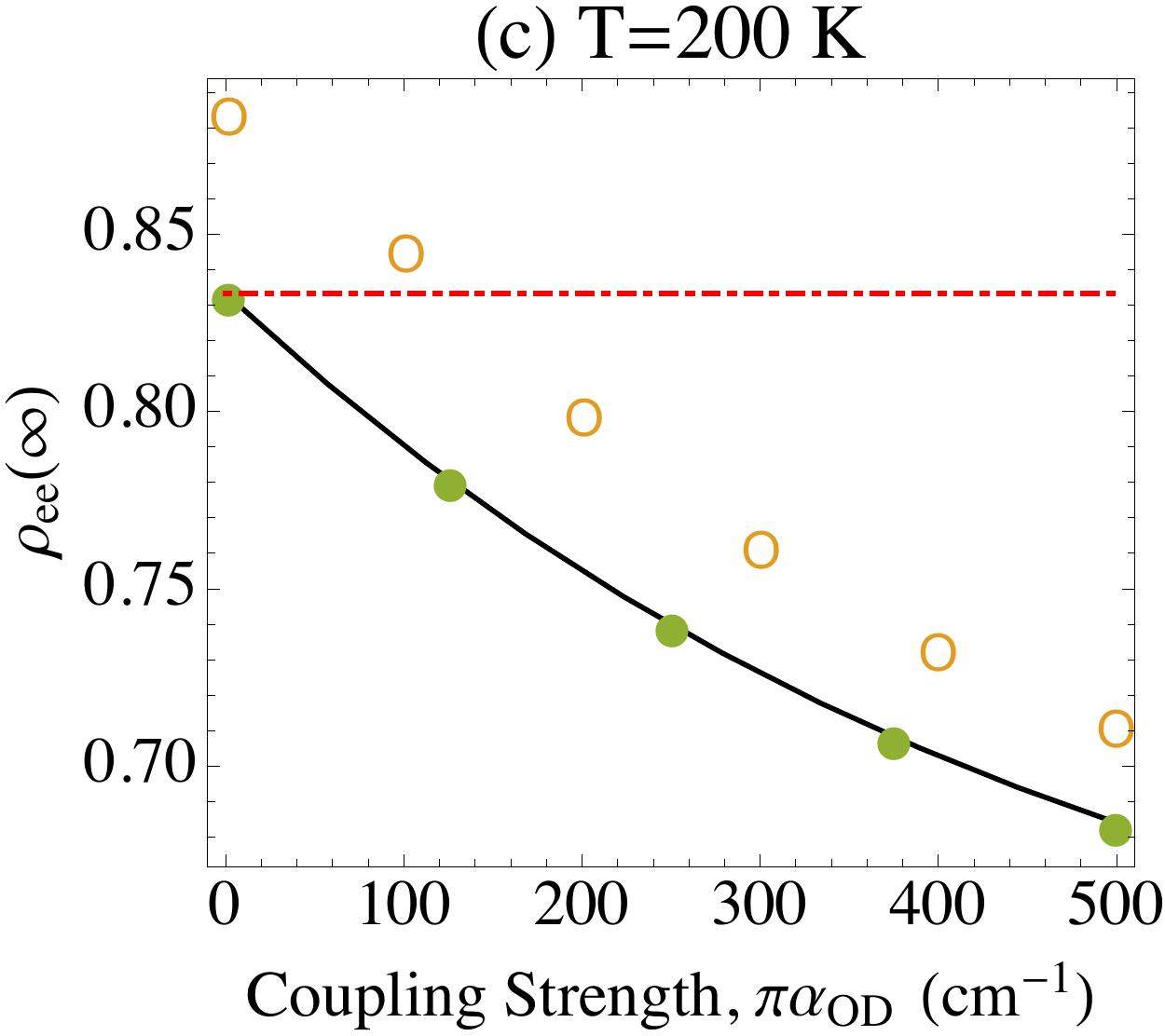}
    \includegraphics[width=0.222\textwidth]{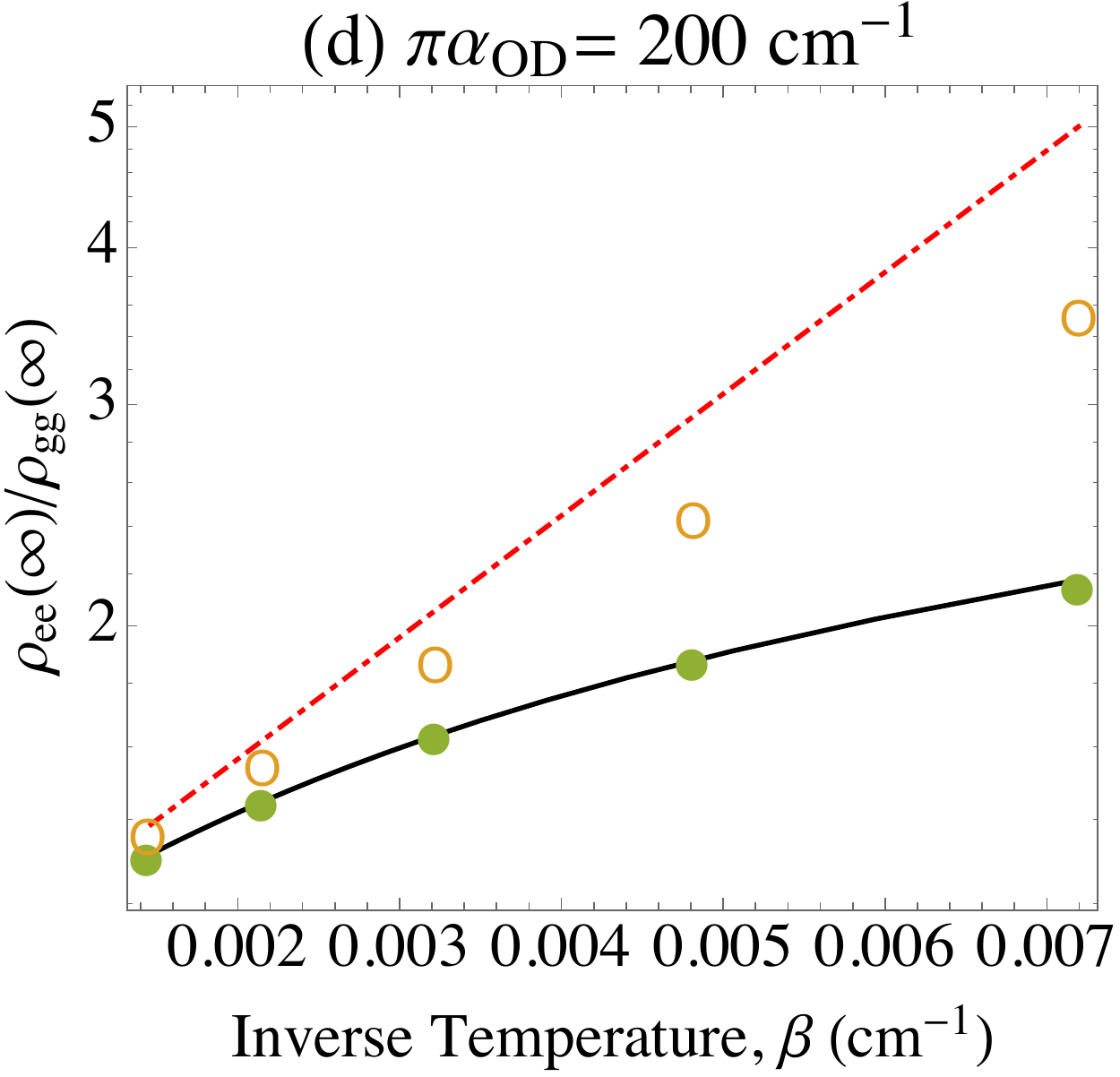}
			\caption[Comparison of the population dynamics predicted by the reaction coordinate master equation, the Zusman equations, and the hierarchical equations of motion for an overdamped spectral density.]{(a,b) Comparison of the dimer site population dynamics $\rho_{11}(t)$ calculated from the RCME (solid curves), Zusman equations (open-points) and the HEOM (solid-points) for the coupling strengths indicated and $T=300$~K. 
			(c) Steady-state population in the dimer eigenstate basis (upper eigenstate, $\rho_{ee}(t)$) as a function of system-environment coupling strength for all three theories and the canonical equilibrium state (dot-dashed).
			(d) Variation of the dimer eigenstate population ratio against inverse  
			temperature for all three theories.
			The other parameters are $\Delta=200$~cm$^{-1}$, $\epsilon=100$~cm$^{-1}$, and $\omega_c=53.08$~cm$^{-1}$} 
			\label{fig:fig2}
		\end{figure}

In Fig.~\ref{fig:fig2} (a) and (b) we compare the short time population dynamics 
of site 1 as predicted by the RCME (solid curves), Zusman equations (open points) and the HEOM (solid points). We consider an overdamped environment and take parameters representative of a subset of the Fenna-Matthews-Olson complex~\cite{ishizaki:234111}. 
We find excellent agreement between the RCME, Zusman equations, and the HEOM 
at both weak and strong coupling to the environment in this regime, capturing the transition from coherent to incoherent energy transfer~\cite{PhysRevLett.103.146404}. 
It is evident 
that 
by including the RC into the system Hamiltonian, we are able to faithfully represent all relevant system-environment correlations in the dimer evolution. Moreover, for this overdamped spectral density, 
the residual environmental influence is sufficiently strong to suppress significant oscillations in the RC degrees of freedom. 
We are therefore in the high friction limit, and 
the Zusman equations are expected to provide a good description of the system dynamics on transient timescales. 

\subsubsection{Non-canonical equilibrium states and the dynamical generation of correlations}

Nevertheless, the semiclassical approximations inherent within the Zusman treatment still manifest themselves on longer timescales, as demonstrated in Fig.~\ref{fig:fig2} (c) and (d). 
Here, we see that they cannot correctly capture the equilibration behaviour of the system in the long-time limit. 
For example, at a temperature of $T=200$~K, the Zusman equations clearly do not lead to the correct steady-state population in the excitonic basis (i.e.~the eigenbasis of $\frac{\epsilon}{2}\sigma_z+\frac{\Delta}{2}\sigma_x$), even for very weak system-environment coupling strengths. 
Interestingly, 
with increasing coupling strength (or decreasing temperature), the steady-states derived from both the RCME and the HEOM deviate noticeably from the canonical thermal state; that is, the state
\begin{equation}\label{eq:cantherm}
\rho_{\rm S_{th}} = \frac{e^{-\beta (\frac{\epsilon}{2} \sigma_{z}+\frac{\Delta}{2}\sigma_x) }}{\mathcal{Z}_C},
\end{equation}
where $\mathcal{Z}_C = \operatorname{tr}_{\rm S}\left\{e^{-\beta (\frac{\epsilon}{2} \sigma_{z}+\frac{\Delta}{2}\sigma_x)}\right\}$.
As shown in Ref.~\onlinecite{PhysRevA.90.032114}, 
this is a consequence of significant and long lasting correlations  accumulated between the system and environment at both weak and strong coupling, with the result that the canonical thermal state no longer describes the true equilibrium state of the system.
Though such steady-state correlations are correctly captured through the RCME, they are not in the Zusman equations. Hence, we find that it is necessary to retain a full quantum treatment of the dimer-RC interaction within the RCME 
to accurately describe our system 
over all timescales.
In fact, the steady-state of the RCME may be compactly expressed 
as a canonical thermal state with respect to the full RC Hamiltonian 
\begin{equation}
\rho(t\rightarrow\infty) = \frac{e^{-\beta \left(\frac{\epsilon}{2} \sigma_{z}+\frac{\Delta}{2}\sigma_x+\lambda\sigma_z\left(a^\dagger+a\right)+\Omega a^\dagger a\right) }}{\mathcal{Z}_{NC}},
\label{eq:non_can}
\end{equation}
where $\mathcal{Z}_{NC} = \operatorname{tr}\left\{e^{-\beta \left(\frac{\epsilon}{2} \sigma_{z}+\frac{\Delta}{2}\sigma_x+\lambda\sigma_z\left(a^\dagger+a\right)+\Omega a^\dagger a\right)}\right\}$. 
On tracing over the RC or TLS, this represents a non-canonical dimer equilibrium state or a non-thermal environmental state, respectively.
The former may be benchmarked against the HEOM, and proves to accurately capture deviations in the steady-state of the dimer due to system-environment correlations~\cite{PhysRevA.90.032114}.
Of course, the influence of correlations is also dependent on the temperature of the bosonic environment. 
Fig.~\ref{fig:fig2} (d) demonstrates that at very large temperatures the steady-states obtained from the RCME, the HEOM, and the Zusman equations begin to agree, converging towards the canonical thermal state.
Here we enter a regime in which the quantum correlations shared between the system and environment are suppressed, such that the semiclassical approximation is adequate to describe the dimer behaviour even at relatively strong system-environment coupling. 

\begin{figure}[t]
\center\hspace{-0.5cm}
\includegraphics[width = 0.237\textwidth]{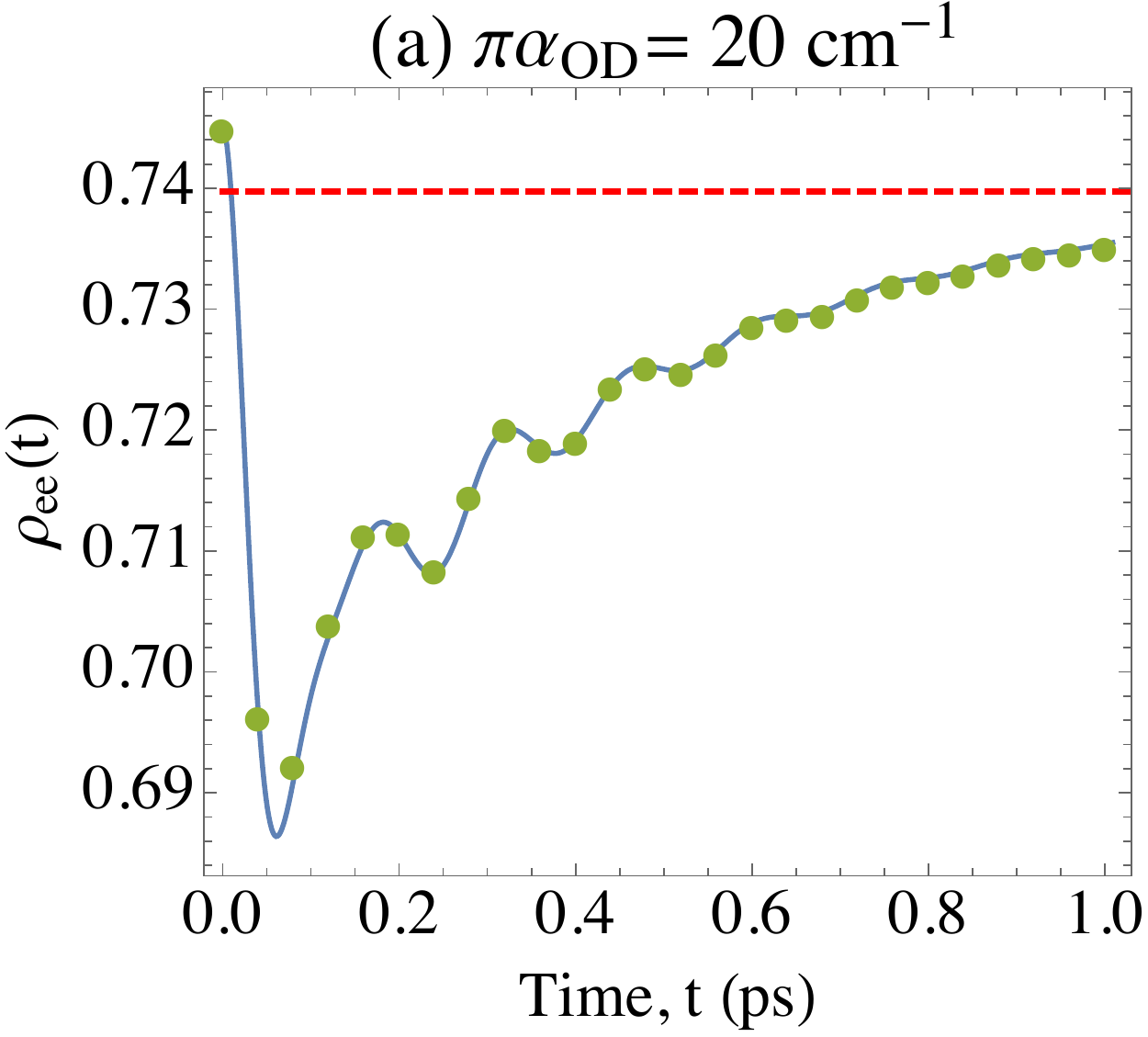}
\includegraphics[width = 0.237\textwidth]{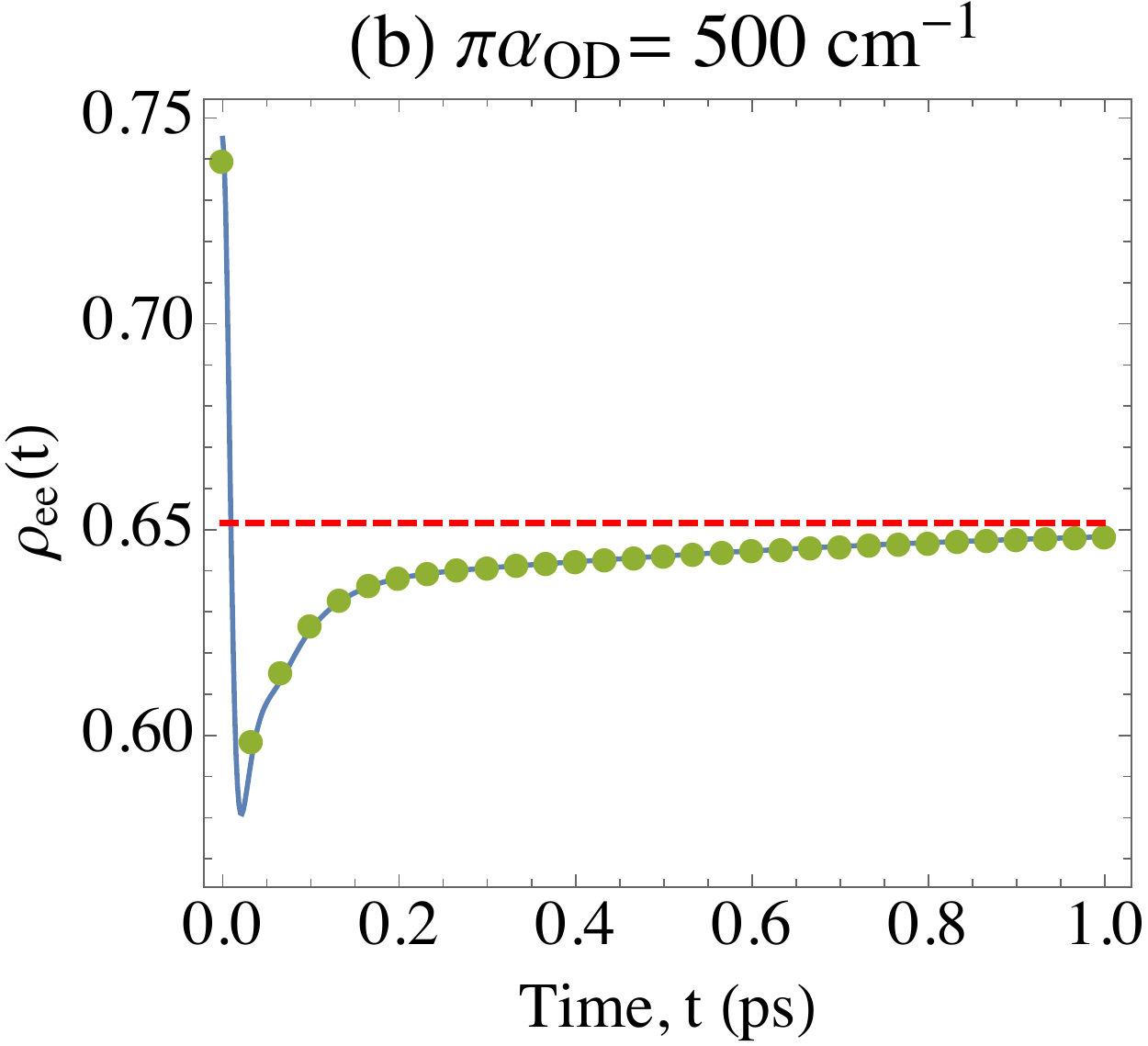}\\
\vspace{0.25cm}\hspace{-0.5cm}
\includegraphics[width = 0.24\textwidth]{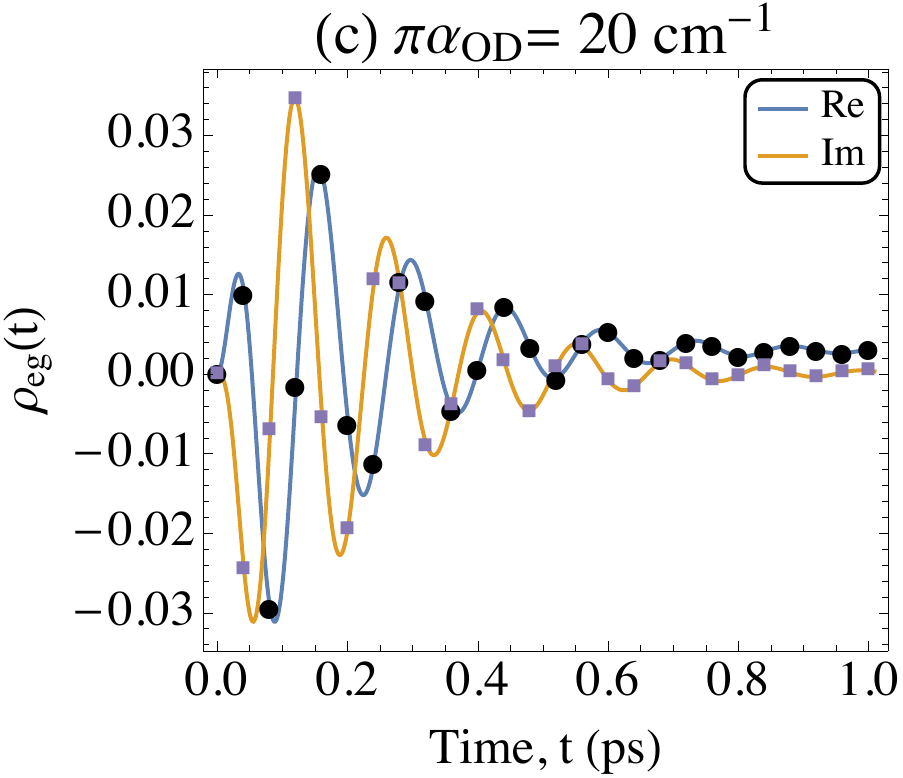}
\includegraphics[width = 0.24\textwidth]{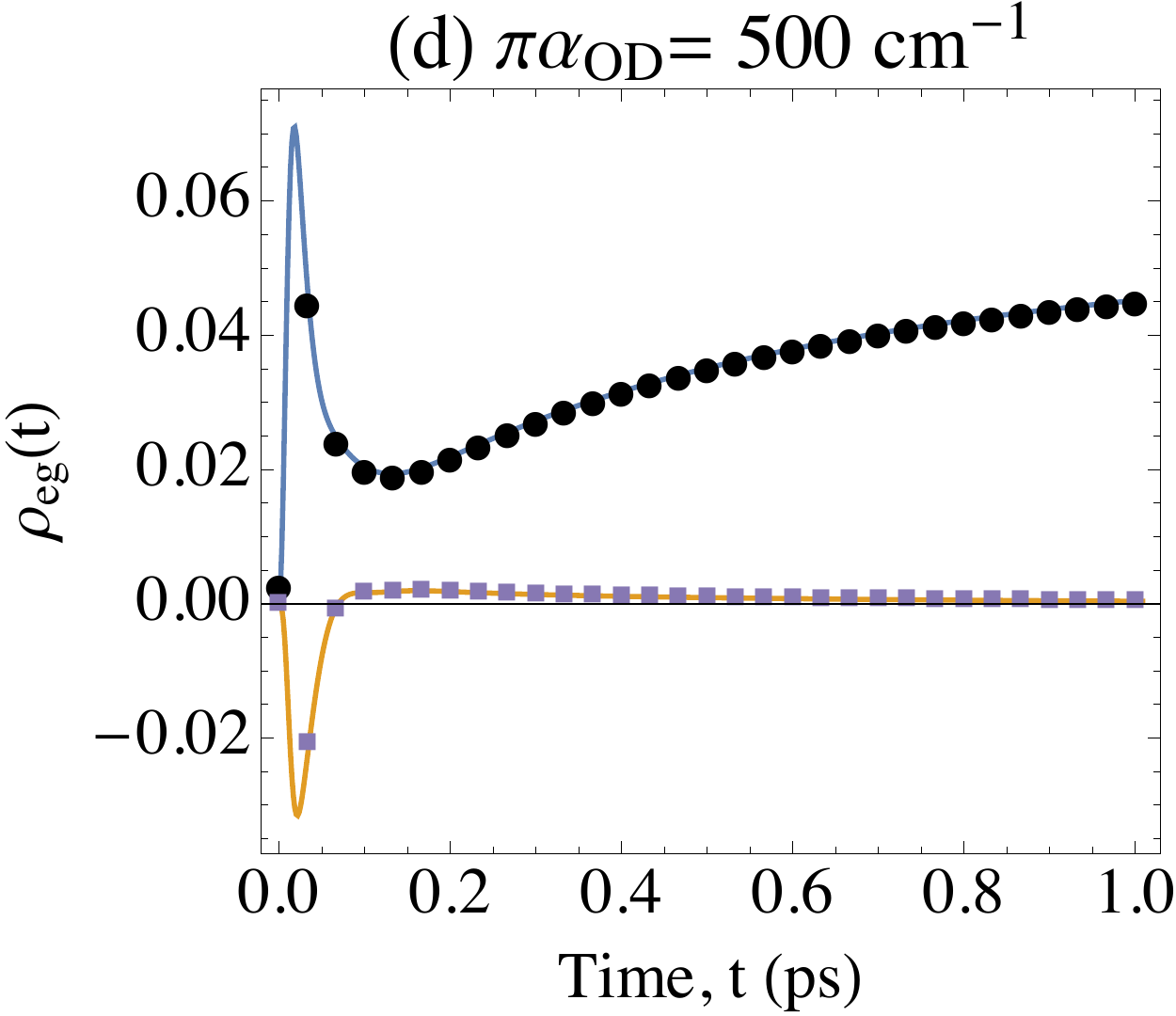}
\caption{Top: Excitonic upper eigenstate population dynamics, $\rho_{ee}(t)$, for a  system initiated in its canonical thermal state given in Eq.~(\ref{eq:cantherm}). The solid curve is calculated with the RCME and the points are obtained from the HEOM. The dashed lines represent the non-canonical system steady-state given by tracing over the RC in Eq.~(\ref{eq:non_can}). Bottom: The corresponding real and imaginary parts of the coherence in the dimer excitonic eigenbasis calculated using the RCME (solid) and the HEOM (points). 
The other parameters are $\omega_c=53.08$~cm$^{-1}$, $\Delta=200$~cm$^{-1}$, $\epsilon=100$~cm$^{-1}$, and $T=300$~K. }
\label{fig:init_therm}
\end{figure}	

System-environment correlations also have important consequences for probing the dynamics in EET systems.
For example, consider a system that is initially in (quasi) equilibrium with its surrounding environment, before being perturbed by an external 
field. 
Na\"ively, one might assume that the initial state of the system in this situation should be the canonical thermal state with respect to the internal system Hamiltonian. 
Our previous arguments, however, demonstrate that this assumption may be misleading in the context of our molecular dimer system, as shown explicitly in Fig.~\ref{fig:init_therm}. Here, we consider the dynamical evolution of the dimer in its excitonic eigenbasis, when the system is initiated in the canonical thermal state given by Eq.~(\ref{eq:cantherm}). 
We see that the subsequent dynamics can display coherent oscillations in the dimer eigenbasis and even the generation of excitonic coherences, before relaxation to an equilibrium state that differs from the initial thermal state. Note that this is true even when the equilibrium state 
is close to the canonical state, as in the left panels of the figure. 

Behaviour of this kind is markedly different to that expected from 
less sophisticated master equation techniques in which the system-environment coupling is treated perturbatively. Often, 
such approaches 
lead to a complete absence of dynamical evolution 
in the dimer eigenbasis for a system initialised in a canonical thermal state, 
since 
Eq.~(\ref{eq:cantherm}) is the expected equilibrium steady-state when the environmental influence is a weak perturbation.

\subsection{Underdamped spectral density}\label{sec:UDdyn}

\begin{figure}[t]
\center
\includegraphics[width=0.45\textwidth]{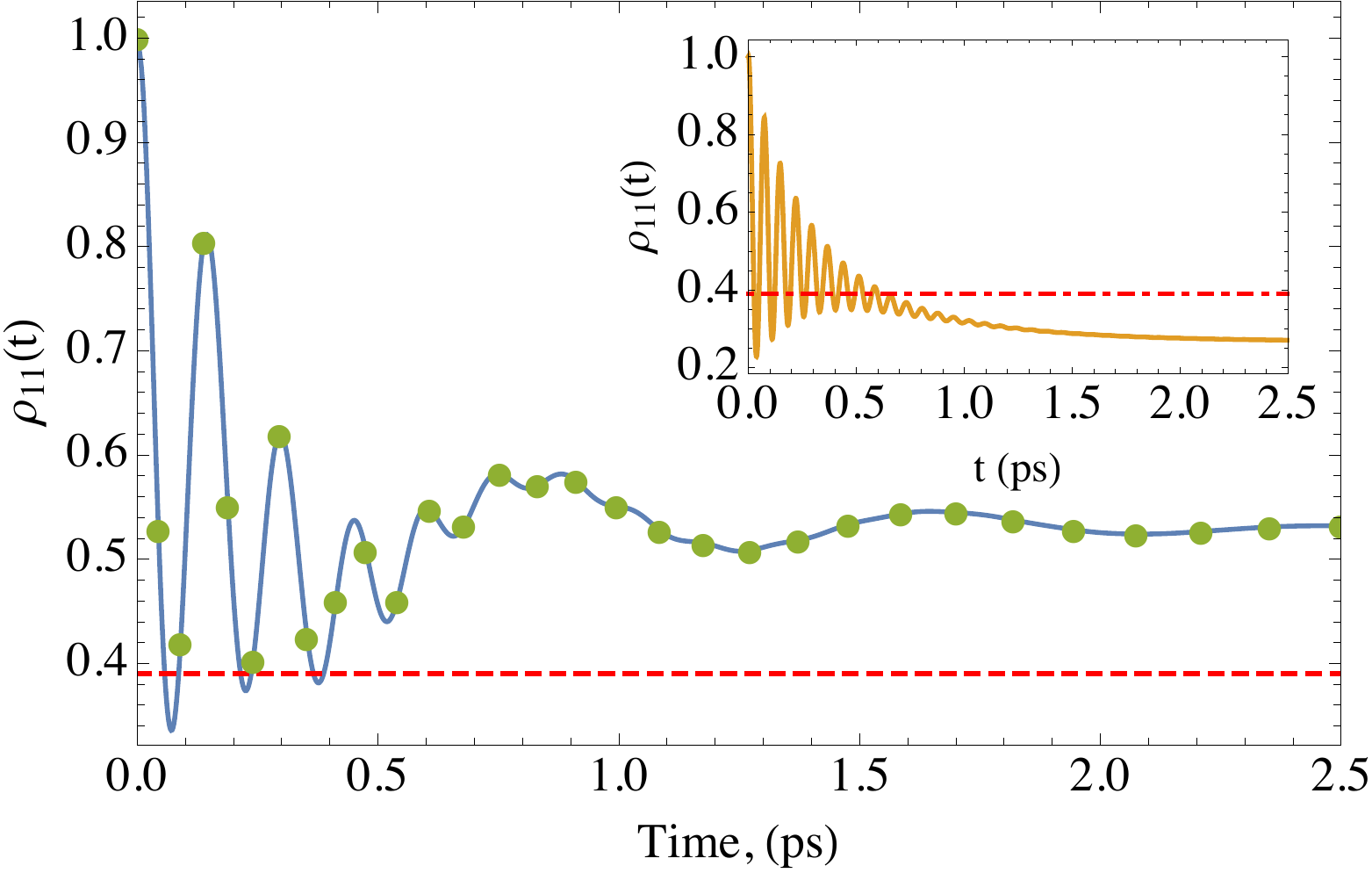}\\
\includegraphics[width=0.45\textwidth]{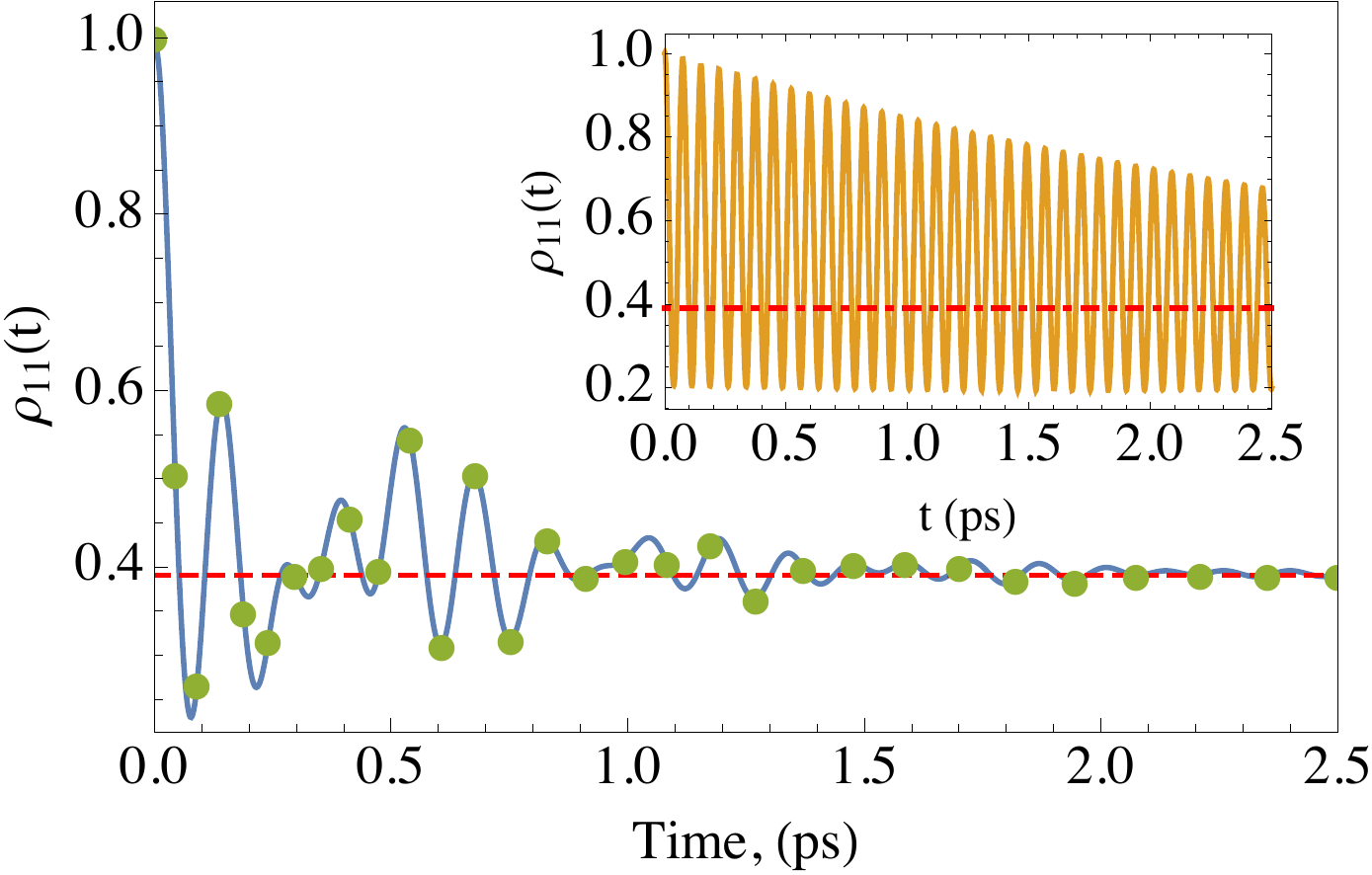}
\caption{Dimer site population dynamics for an underdamped spectral density with (top) $\omega_0=40$~cm$^{-1}$, $\pi\alpha_{\rm UD}=20$~cm$^{-1}$, and (bottom) $\omega_0=220$~cm$^{-1}$, $\pi\alpha_{\rm UD}=10$~cm$^{-1}$. In the main plots we compare the RCME (solid curves) and HEOM (points), while results from the Zusman equations are shown in the insets. Dashed lines denote the non-canonical steady-state values (reached on a very slow timescale for the top plot). The other parameters are $\Delta=200$~cm$^{-1}$, $\epsilon=100$~cm$^{-1}$, $\Gamma=10$~cm$^{-1}$, and $T=300$~K.}
\label{fig:UDp1}
\end{figure}

We shall now move on to discuss the impact of an underdamped spectral density on the EET dynamics of our dimer. The resultant complex system dynamics have particular relevance to EET in the presence of structured environments, where the system may be strongly coupled to specific lossy modes that dominate regions of the pigment-protein vibrational spectrum.  

\begin{figure}[t!]
\center
\includegraphics[width=0.45\textwidth]{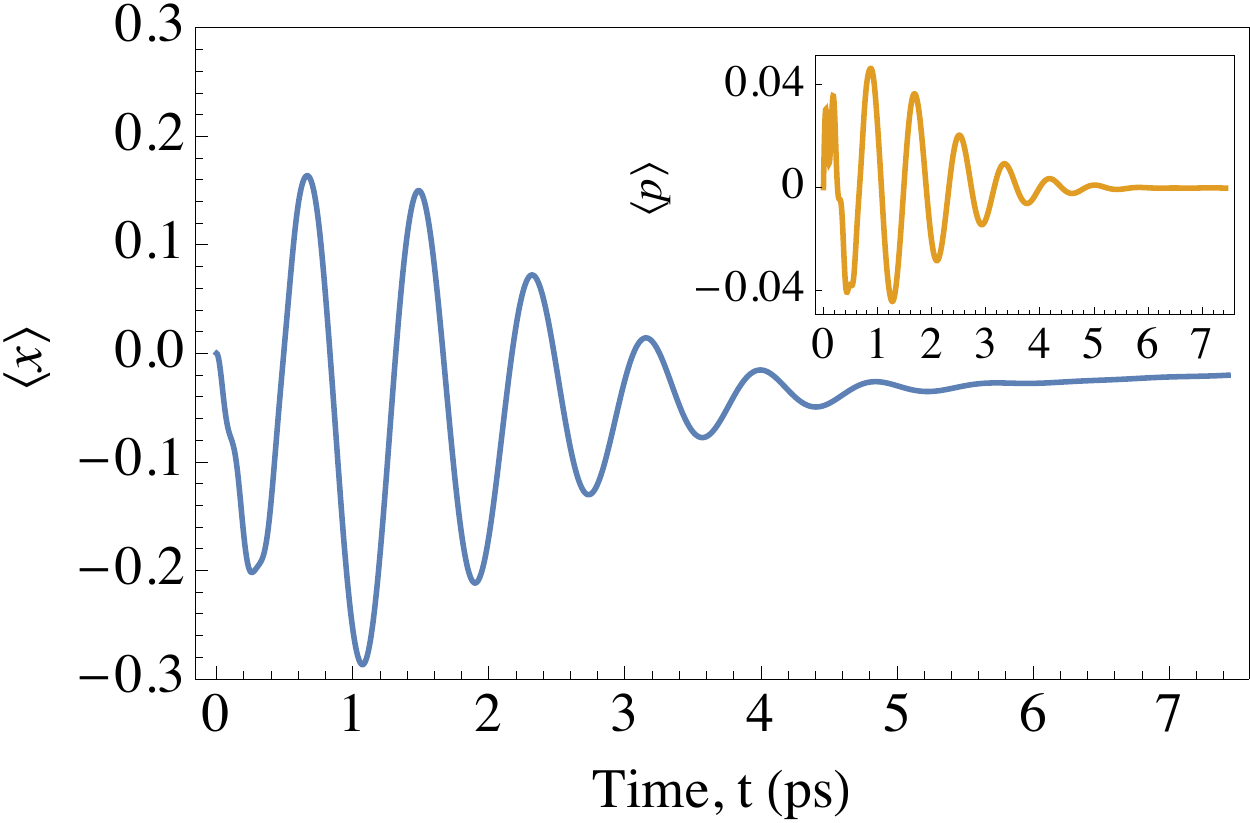}\\
\includegraphics[width=0.45\textwidth]{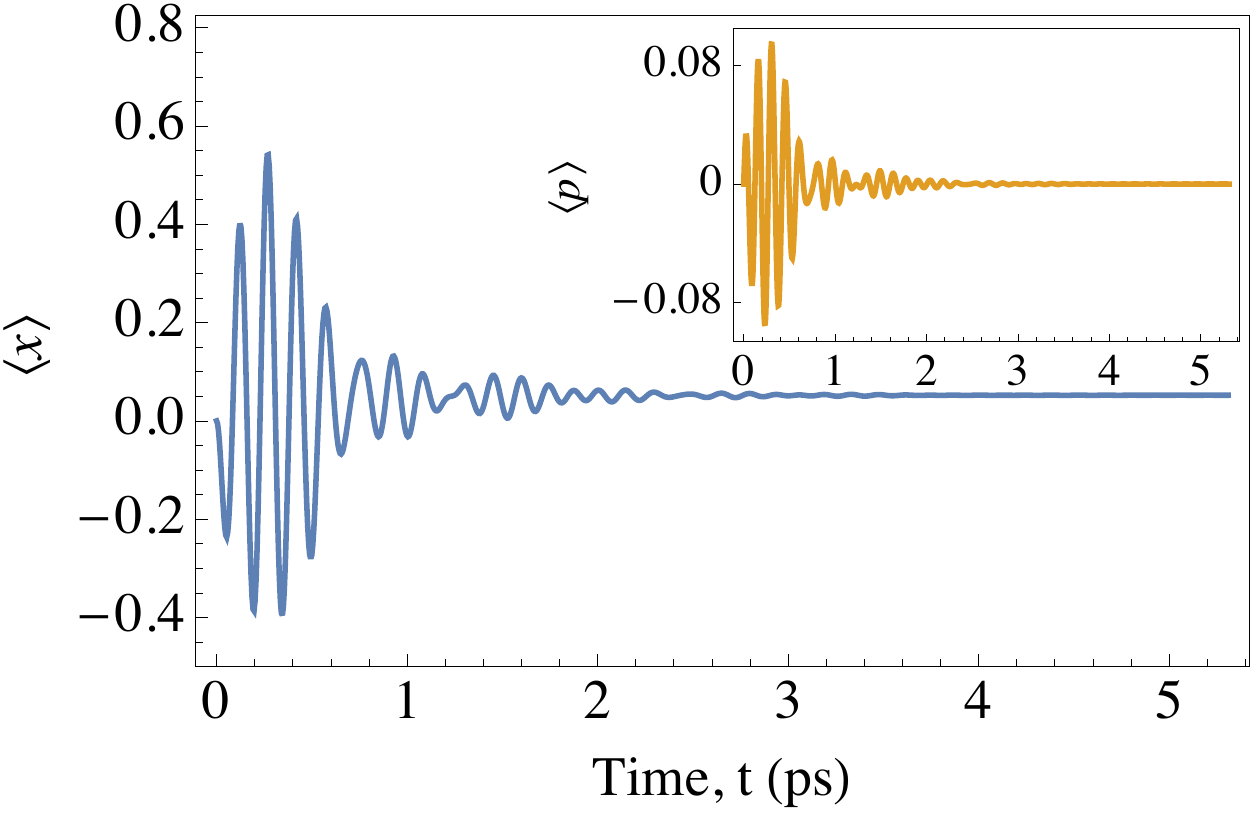}
\caption{The RC position (main) and momentum (inset). Top: $\omega_0=40$~cm$^{-1}$ and $\pi\alpha_{\rm UD}=20$~cm$^{-1}$. Bottom: $\omega_0=220$~cm$^{-1}$ and $\pi\alpha_{\rm UD}=10$~cm$^{-1}$. Other parameters are the same as in Fig.~\ref{fig:UDp1}.}
\label{fig:UDRC}
\end{figure}

In the main plots of Fig.~\ref{fig:UDp1} we compare the population and coherence dynamics obtained from the RCME (solid curves) to the HEOM (points) for both a low 
and high frequency underdamped spectral density, where Zusman predictions are shown in the inset and non-canonical steady-state values are also indicated (dashed lines). 
In both cases we see that the Zusman equations completely fail to capture the correct behaviour, which 
may be attributed to the dynamical response of the RC itself. Considering, for example, the low frequency spectral density, 
when one compares the RC frequency ($\Omega=40$~cm$^{-1}$) to the friction acting on the mode ($\eta\approx5$~cm$^{-1}$, 
obtained from the mapping) 
we find that the RC is weakly damped. Hence, the limits taken to derive the Zusman equations are invalid, 
i.e.~the momentum of the RC does not remain in thermal equilibrium in this regime. 
This is shown explicitly in Fig.~\ref{fig:UDRC}, where we plot the dynamical evolution 
of the RC position $\langle{\hat{x}(t)}\rangle=\tr\{\hat{x}\rho(t)\}$, and momentum $\langle{\hat{p}(t)}\rangle=\tr\{\hat{p}\rho(t)\}$, defined as
\begin{equation}
\hat{x}=\sqrt{\frac{1}{2\Omega}}(a^\dagger+a)\hspace{0.5cm}\text{and}\hspace{0.5cm}\hat{p}=i\sqrt{\frac{\Omega}{2}}(a^\dagger-a).
\end{equation}
We see pronounced oscillations at several frequencies in both the RC position and momentum, which are gradually damped at long times, equilibrating to non-zero values consistent with the 
state given in Eq.~(\ref{eq:non_can}).

The disagreement between the Zusman equations and the RC model is further exacerbated when the underdamped spectrum is tuned close to the dimer resonance, $\zeta=\sqrt{\epsilon^2 +\Delta^2}\approx223$~cm$^{-1}$. 
As can be seen in Fig.~\ref{fig:UDRC}~(b), the RC now undergoes even larger amplitude oscillations. 
In this case the Zusman equations are completely unable to capture the dynamical behaviour of the system, predicting only rapid oscillations in the dimer population. The RCME, on the other hand, shows excellent agreement with the HEOM for both short and long times, capturing perfectly the population dynamics of the dimer. 
In both the low and high frequency cases, the dimer dynamics displays complex beating behaviour, with multiple oscillation frequencies. From the RC formalism we have a clear interpretation for this behaviour in terms of the strong oscillations experienced by the collective coordinate of the environment, 
which in turn leads to modulations of the TLS 
dynamics. 

\begin{figure}[t]
 \center
\hspace{-0.5cm}
\includegraphics[width = 0.24\textwidth]{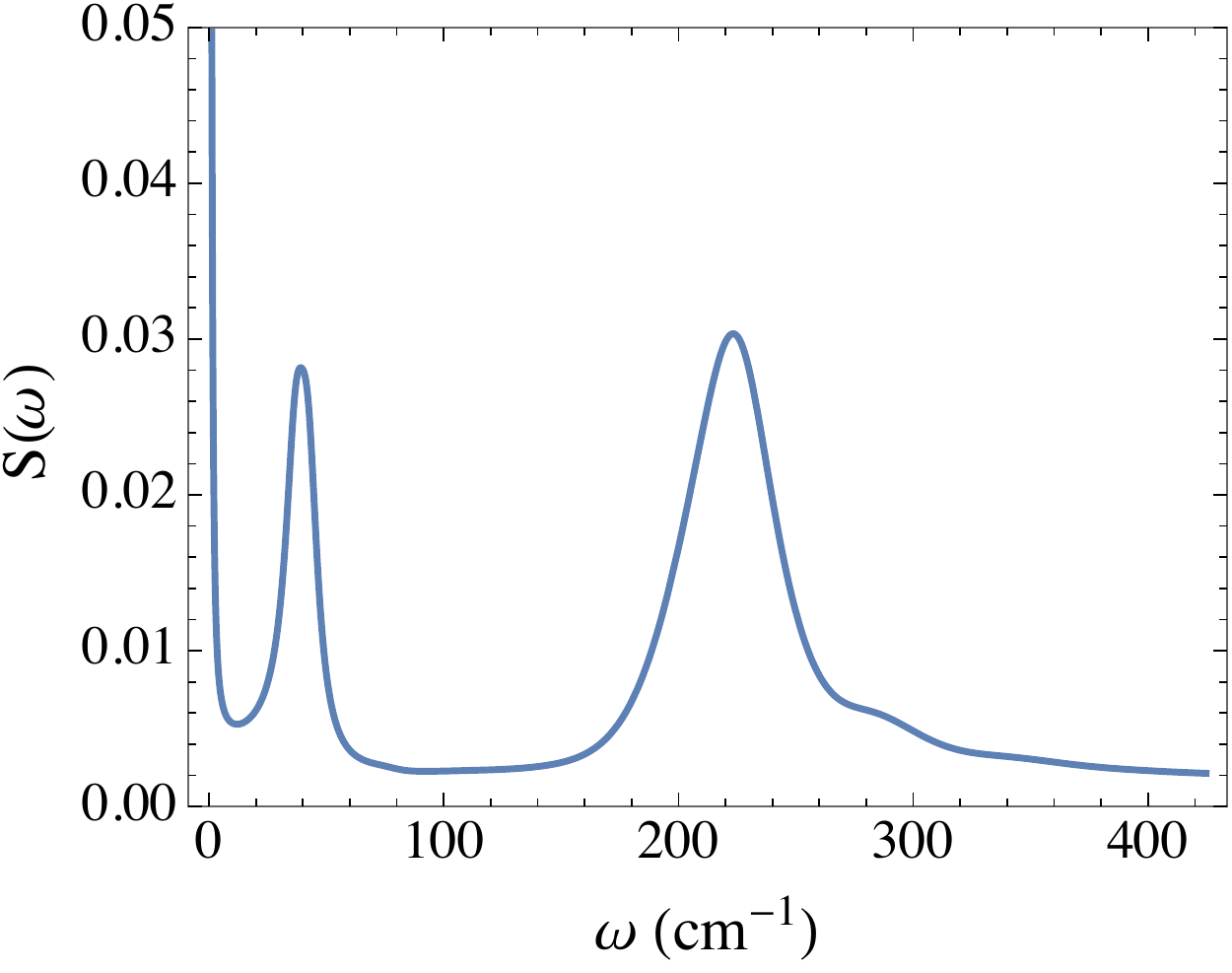}
\includegraphics[width = 0.24\textwidth]{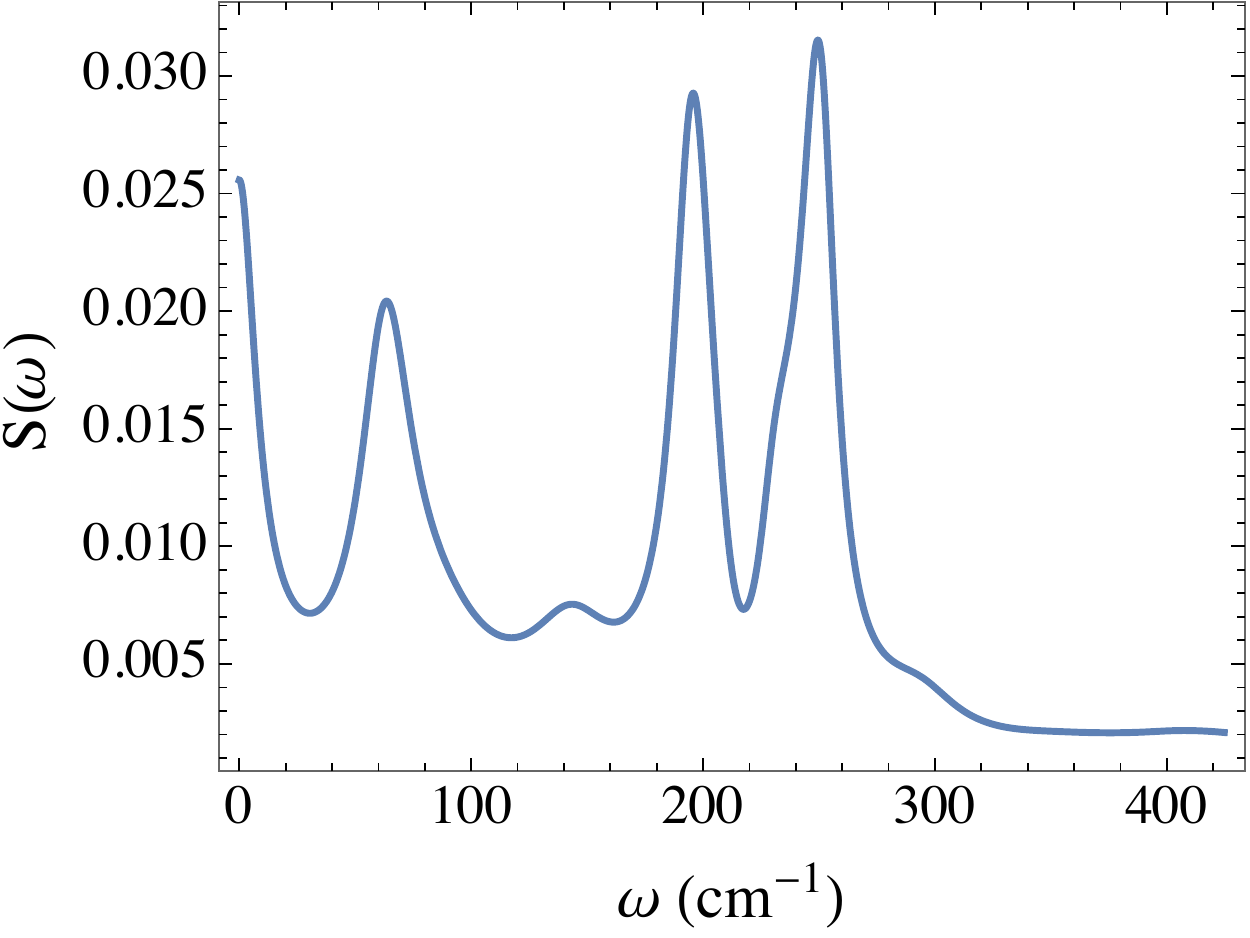}
 \caption[Fourier transforms of the population dynamics.]{Fourier transform of the RCME population dynamics in Fig.~\ref{fig:UDp1}, demonstrating the presence of multiple oscillation frequencies. Left: $\omega_0=40$~cm$^{-1}$. Right: $\omega_0=220$~cm$^{-1}$.}
 \label{fig:Four}
 \end{figure}

This is shown 
explicitly in Fig~\ref{fig:Four}, where we have taken the Fourier transform of the population dynamics given in Fig.~\ref{fig:UDp1}, that is
 \begin{equation}
 S(\omega)=\operatorname{Re}\left[\int\limits_0^\infty dt e^{i\omega t}\left(\rho_{11}(t)-\rho_{11}(t\rightarrow\infty)\right)\right].
 \end{equation}
Here, we have subtracted the steady-state population, $\rho_{11}(t\rightarrow\infty)$, to remove a $\delta$-function contribution. 
For the lower frequency underdamped environment (left plot) we see the presence of two specific frequencies in the spectrum, one at the RC mode frequency ($\omega_0 = 40$~cm$^{-1}$), and the other at the dimer splitting ($\zeta\approx223$~cm$^{-1}$). 
When the characteristic frequency of the spectral density approaches the dimer splitting (right plot) we 
see 
further structure in the oscillation spectrum, 
with the emergence of additional peaks. 
Again, we may explain this by appealing to the physical intuition gained from the RC model. In this case, the environmental response, and thus RC splitting, lies close to the resonant frequency of the dimer, leading to an effective enhancement of the interaction between the dimer and RC. 
As a result of the enhanced coupling, the spectrum of the system cannot be associated to the bare frequencies of the dimer and RC, but rather to the eigenstates of the composite system. 
Hence, we see a double peak structure about $\omega\sim 220$~cm$^{-1}$, split by the RC-dimer coupling strength $2\lambda=66$~cm$^{-1}$. 
This is reminiscent of the vacuum Rabi splitting observed in cavity QED systems, in which the eigenstates of the system are the light-matter entangled dressed states~\cite{steck2007quantum}. 

The presence of multiple oscillation frequencies in the population dynamics has important implications for a number of ongoing experiments on EET systems. 
In particular, the 
discussion 
above demonstrates that it is not 
straightforward to assign electronic and vibrational frequencies in situations where underdamped modes 
are present, as one must also account for the coupling between the molecular dimer and any such 
modes, 
which leads to the formation of vibronic states.

\begin{figure}[t]
 \center
 \includegraphics[width = 0.237\textwidth]{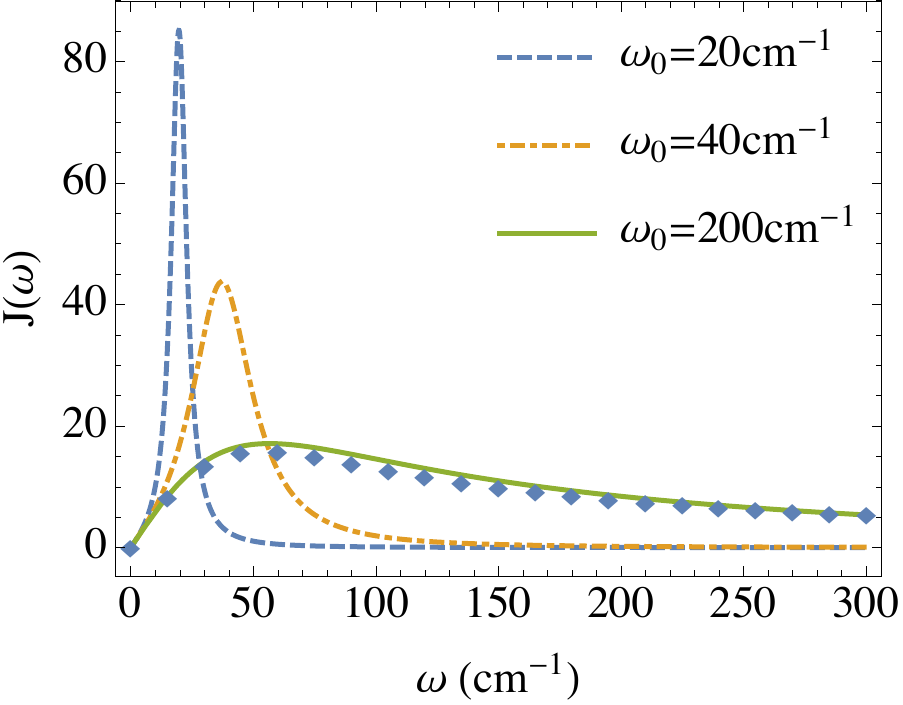}
\includegraphics[width = 0.237\textwidth]{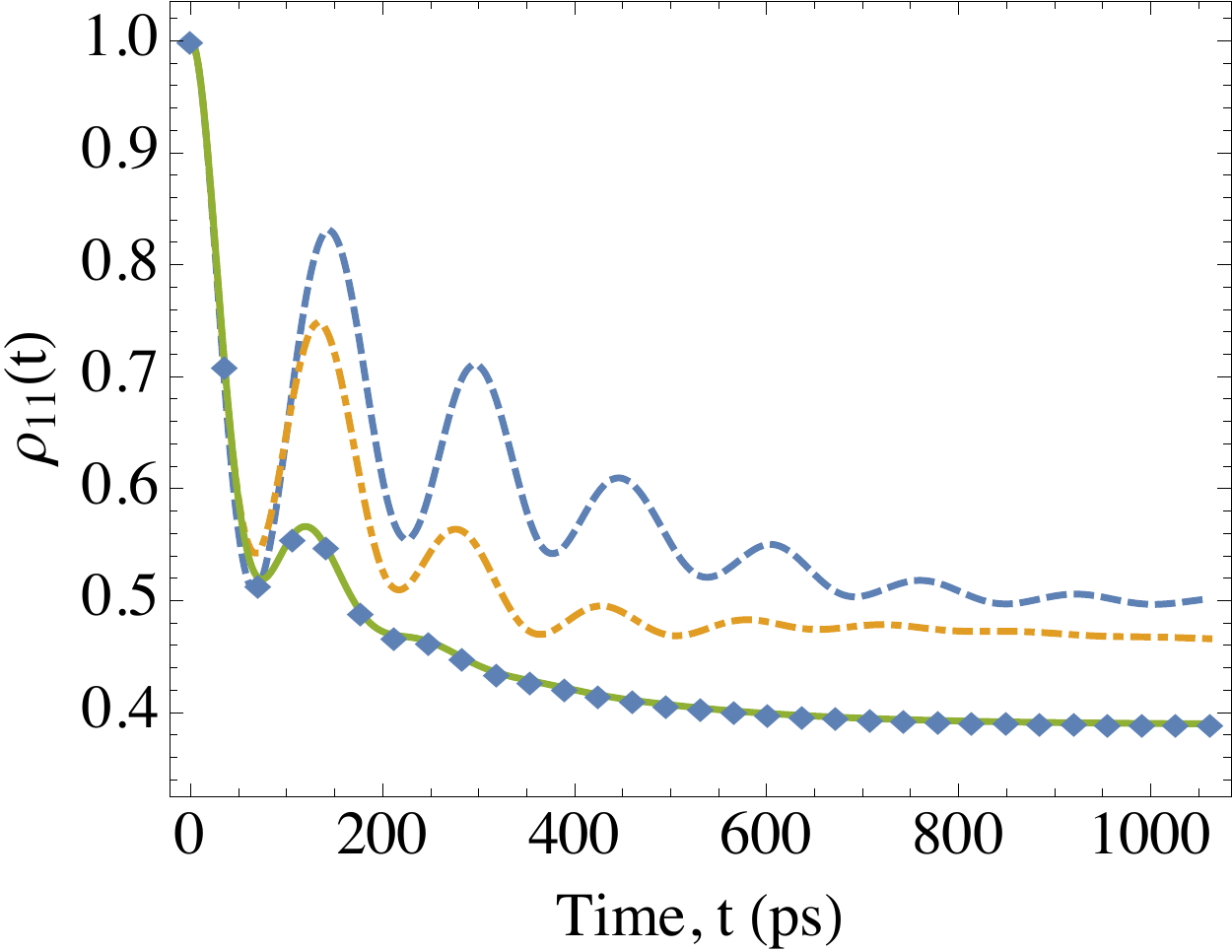}
 \caption[Transition of the underdamped reaction coordinate to the overdamped case.]{Left: Underdamped (curves) and overdamped (points) spectral densities for several environmental frequencies. Right: The population dynamics associated to these spectral densities. Here, we see a smooth transition from the underdamped (curves) to overdamped (points) regime for increasing $\omega_0$. 
The parameters are the same as Fig.~\ref{fig:UDp1}, with $\Gamma=\omega_0^2\omega_c^{-1}$, and $\pi\alpha =100$~cm$^{-1}$. }
 \label{fig:UDtoOD}
 \end{figure}

As a final remark, we can show that the RC model for an underdamped spectrum convergences to the overdamped case in regimes where $\Gamma,\omega_0\gg 1$. 
We do this by defining a cut-off frequency in the underdamped spectrum as $\omega_c = \omega_0^2/\Gamma$, which sets the energy scale of the overdamped spectrum in the appropriate limit. 
Using this definition we fix $\Gamma = \omega_0^2 /\omega_c$ and consider the underdamped spectrum for increasing $\omega_0$ as demonstrated in Fig.~\ref{fig:UDtoOD}.
Here we see a smooth transition between the underdamped and overdamped regimes, with the two agreeing well at large $\omega_0$.

\section{Structured environments}\label{sec:sec5}

\begin{figure}[t]
\center
\includegraphics[width=0.235\textwidth]{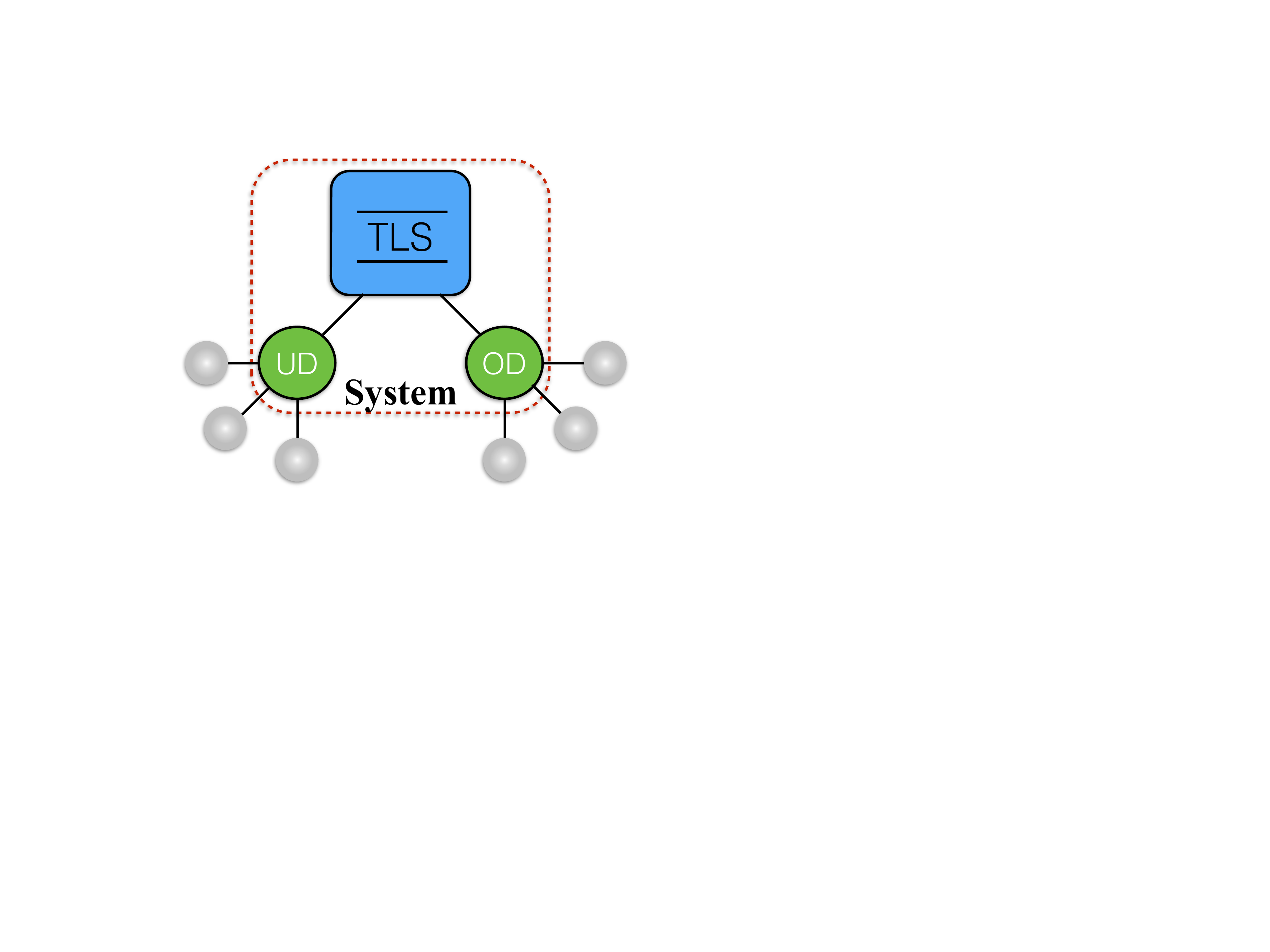}
\includegraphics[width=0.235\textwidth]{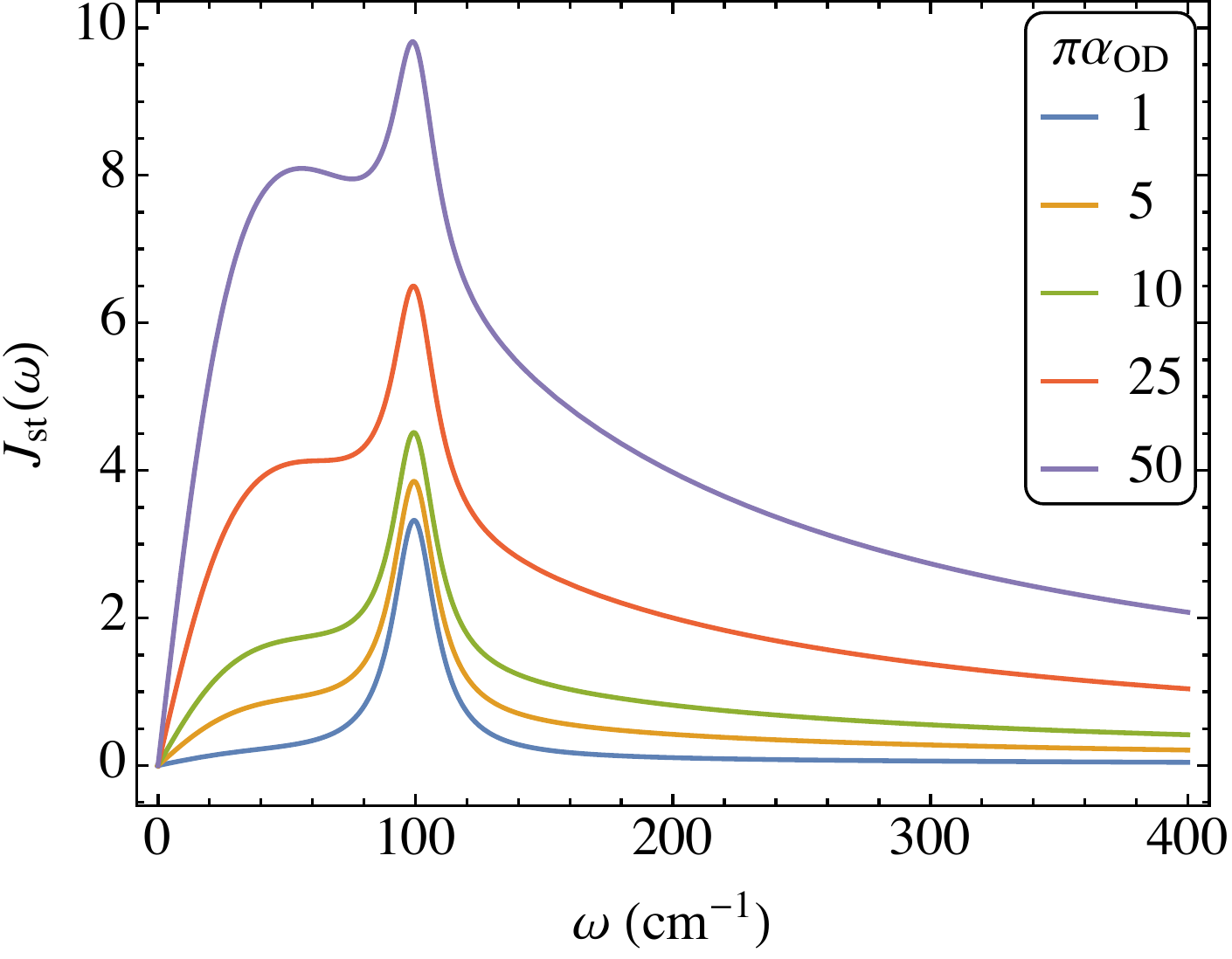}
\caption{Left: Schematic of the RC model for a structured environment with underdamped (UD) and overdamped (OD) components, each coupled to their own residual environment. Right: The structured spectral density at various reorganisation energies of the overdamped contribution. 
The underdamped RC has frequency $\omega_0=100$~cm$^{-1}$, with $\Gamma = 20$~cm$^{-1}$ and $\pi\alpha_{\rm UD} = 2$~cm$^{-1}$.}
\label{fig:struct_schem}
\end{figure}

Having established the validity and potential for 
physical insight 
of 
the RCME when applied 
to overdamped and underdamped environments separately, 
in this section we explore how a structured environment 
impacts upon the energy transfer rate in a molecular 
system.
To do this in a 
consistent manner, we shall consider the dynamics of a dimer coupled to a broad background environment described by an overdamped spectral density, with structure incorporated via 
a second underdamped environment with peak centred around $\omega_0$.
We shall model both environments using the RC mapping outlined previously, extracting an independent RC for each, allowing us to rigorously account for dissipation on the underdamped mode. Furthermore, our enlarged system (see left plot of Fig.~\ref{fig:struct_schem}) naturally captures the vibronic nature imparted on the dimer by {\it both} the underdamped and overdamped components of the environmental spectrum.

The Hamiltonian describing the system and environments may be written as 
\begin{equation}
H_{\rm ST}=H_{\rm D} + \sigma_z\sum\limits_{i=1}^2\sum\limits_k f_k^{(i)}(c_{i,k}+c_{i,k}^\dagger) + H_{\rm B},
\end{equation}
with $H_{\rm B}=\sum\limits_k\omega_k^{(i)}c^\dagger_{i,k}c_{i,k}$ and $H_{\rm D}=\frac{\epsilon}{2}\sigma_z + \frac{\Delta}{2}\sigma_x$. 
The two  
environments are characterised by the spectral densities
\begin{equation}
{J}_i(\omega) = \sum\limits_k\vert f_k^{(i)}\vert^2\delta(\omega-\omega_k^{(i)}),
\end{equation}
with $J_1(\omega)=J_{\rm OD}(\omega)$ and $J_2(\omega)=J_{\rm UD}(\omega)$.
The combination of these terms leads to an effective spectral density with a broad background, given by the overdamped component, and a sharp peak associated to the underdamped contribution. 
Illustrative examples 
are given in the right hand plot of Fig.~\ref{fig:struct_schem}. 

We shall assume that the two 
environments are initially uncorrelated with one another (e.g.~in a thermal state), but are 
able to generate correlations through interactions mediated by the dimer.
This allows each environment to be mapped to the RC model independently. 
Applying the mapping we obtain the system Hamiltonian
\begin{equation}
H_{\rm S}= \frac{\epsilon}{2}\sigma_z + \frac{\Delta}{2}\sigma_x + \sigma_z \sum\limits_i\lambda_{i} \left(a_i^\dagger+a_i\right) + \sum\limits_i\Omega_ia^\dagger_ia_i,
\end{equation}
with $a_1$ ($a_2$) the annihilation operator of the RC associated with the underdamped (overdamped) environment. 
Each RC then couples to an independent residual environment, giving the interaction Hamiltonian 
\begin{equation}
H_{\rm I}=\sum\limits_i \left(a^\dagger_i  + a_i\right) \sum\limits_k g_k^{(i)}\left(b_{i,k}^\dagger + b_{i,k}\right).
\end{equation}
Here, $b_{i,k}$ ($b_{i,k}^\dagger$) is the annihilation (creation) operator for the $k^{th}$ mode of each residual environment ($i=1,2$), which are characterised by the spectral densities $\tilde{J}_{i}(\omega)=\sum_k \vert g_k^{(i)}\vert^2\delta(\omega-\omega_k^{(i)})=\gamma_i \omega$.
The parameters describing the RCs can then be found in terms of the original spectral densities using the relations given in Section~\ref{sec:sec2}.

By following the RCME derivation for each independent environment we obtain an equation of motion describing the dynamics of the dimer TLS and {\it both} RCs
\begin{align}\label{eq:full}
\frac{\partial\rho(t)}{\partial t}=-i\left[H_{\rm S},\rho(t)\right]-\sum\limits_i& \Big(\left[\hat{A}_i,\left[\hat{\chi}_i,\rho(t)\right]\right]\nonumber\\
&+\left[\hat{A}_i,\left\{\hat\Xi_i,\rho(t)\right\}\right]\Big),
\end{align}
with $\hat{A}_{i} = (a^\dagger _i+ a_i)$, and the rate operators $\hat\chi_i$ and $\hat{\Xi}_i$ defined in analogy to the single RC case in Eqs.~(\ref{eqn:Xi}) and~(\ref{eqn:chi}). 

\begin{figure}[t]
 \center
 \includegraphics[width=0.4\textwidth]{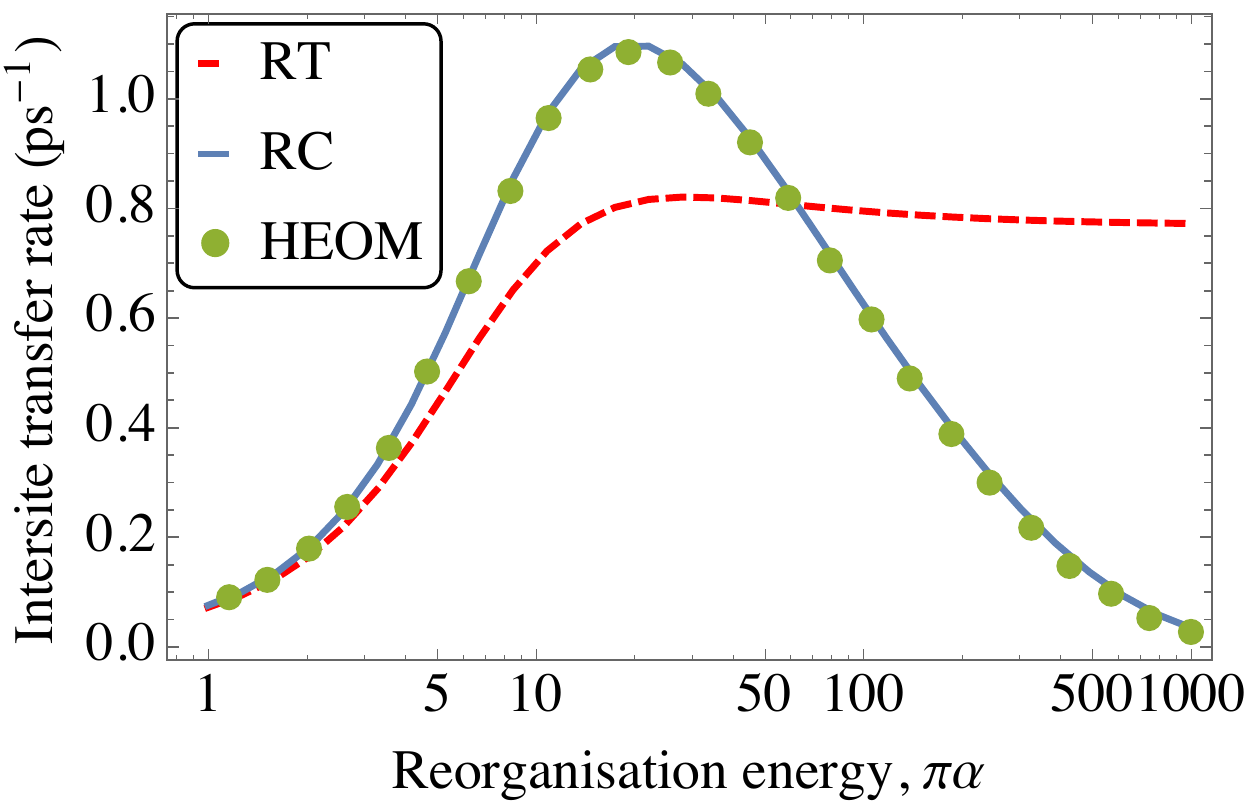}\\
  \includegraphics[width=0.4\textwidth]{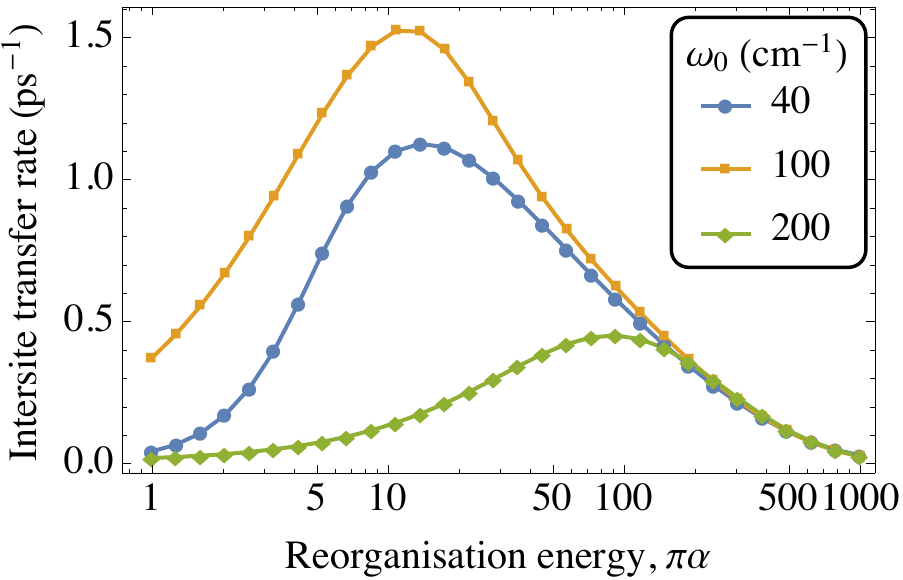}
			\caption[Comparing the energy transfer rate predicted by the RC, Redfield, and hierarchy.]{Variation in dimer inter-site energy transfer rate $k_{1\rightarrow2}$ for increasing reorganisation energy. Top: Predictions from the RCME (solid curve), the HEOM (points) and weak-coupling Redfield theory (RT, dashed curve) for a single overdamped environment. Both the RCME and HEOM predict a smooth peak in the rate at intermediate reorganisation energy, 
whereas the Redfield master equation fails to capture the correct behaviour for all but the weakest coupling strengths. Bottom: RCME predictions for a single underdamped environment for several characteristic frequencies $\omega_0$, with $\Gamma = 20$~cm$^{-1}$. Notice that the largest transfer rates occur for $\omega_0=100$~cm$^{-1}$, which is close to the resonance of the dimer TLS. Other parameters are $\epsilon=100$~cm$^{-1}$, $\Delta=40$~cm$^{-1}$, $\omega_c=53$~cm$^{-1}$, and $T=300$~K.} 
			\label{fig:rate}
\end{figure}

We shall now focus on a quantitative analysis and comparison of the dimer EET rate for both structured and unstructured environments. 
In regimes for which the dimer splitting is greater than the tunnelling rate (in this case $\epsilon = 100$~cm$^{-1}$ and $\Delta = 40$~cm$^{-1}$), we can do so by defining a rate using the
the classical  
equations~\cite{ishizaki:234111}
\begin{equation}
\begin{split}
&\frac{dP_{1}(t)}{d t}=-k_{1\rightarrow2}P_{1}+k_{2\rightarrow1}P_{2},\\
&\frac{dP_{2}(t)}{d t}=k_{1\rightarrow2}P_{1}-k_{2\rightarrow1}P_{2}.
\end{split}
\end{equation}
Here, $P_1$  ($P_2$) is the population at dimer site $1$ ($2$) and $k_{1\rightarrow 2}$ ($k_{2\rightarrow 1}$) is the transfer rate between sites $1$ and $2$ ($2$ and $1$). 
These equations lead to purely exponential decays of the dimer populations, thus neglecting all coherent contributions in the energy transfer dynamics.
Though a coarse approximation, this procedure gives insight into the overall transfer rate, and is accurate in regimes where the tunnelling between sites is weak and coherent dynamics is consequently suppressed.
 
We first consider the case of a single overdamped environment~\cite{ishizaki:234111}. In Fig.~\ref{fig:rate} (top) we plot the inter-site transfer rate as a function of the reorganisation energy calculated from the RCME (solid curve), the HEOM (points), and a Redfield master equation (dashed curve) in which the system-environment coupling is treated perturbatively~\cite{breuer2007theory}. We see that the RCME perfectly captures the smooth peak in the rate predicted by the HEOM as the reorganisation energy is increased. 
As has previously been shown~\cite{ishizaki:234111,varET}, Redfield theory fails even to qualitatively
 capture this behaviour, 
 plateauing at large reorganisation energies.
We may also explore the transfer rate in the presence of a single underdamped environment using the RCME. 
As shown in Fig.~\ref{fig:rate} (bottom), much like the overdamped example, the EET rate in the underdamped case shows a peak at some intermediate coupling strength, the position and height of which is highly dependent on the characteristic frequency of the environment. 
Specifically, the transfer rate reaches its maximum when the peak of the underdamped spectrum approaches the resonance of the dimer ($\zeta\approx 108$~cm$^{-1}$ here). 
As was discussed in Sec.~\ref{sec:UDdyn}, when the dimer and the RC are close to resonant, the vibronic states of the composite system play a significant role in the system dynamics. In this case they act to increase the number of pathways available for energy to be transferred between the two sites of the dimer~\cite{killoran2015}, thus enhancing the EET rate. 
   \begin{figure}[t]
 \center
 \includegraphics[width=0.475\textwidth]{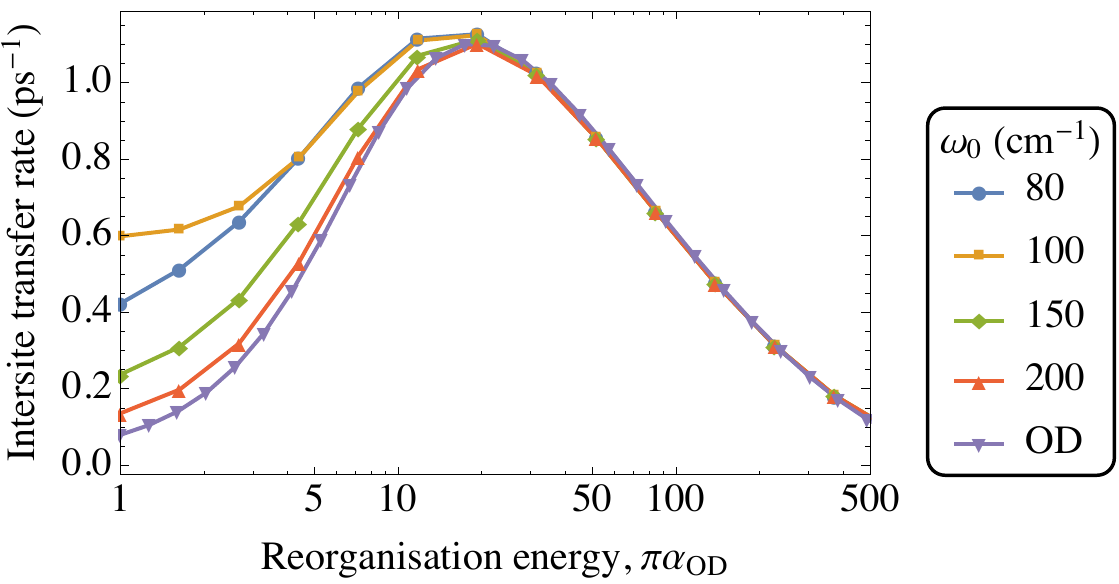}\\
  \includegraphics[width=0.475\textwidth]{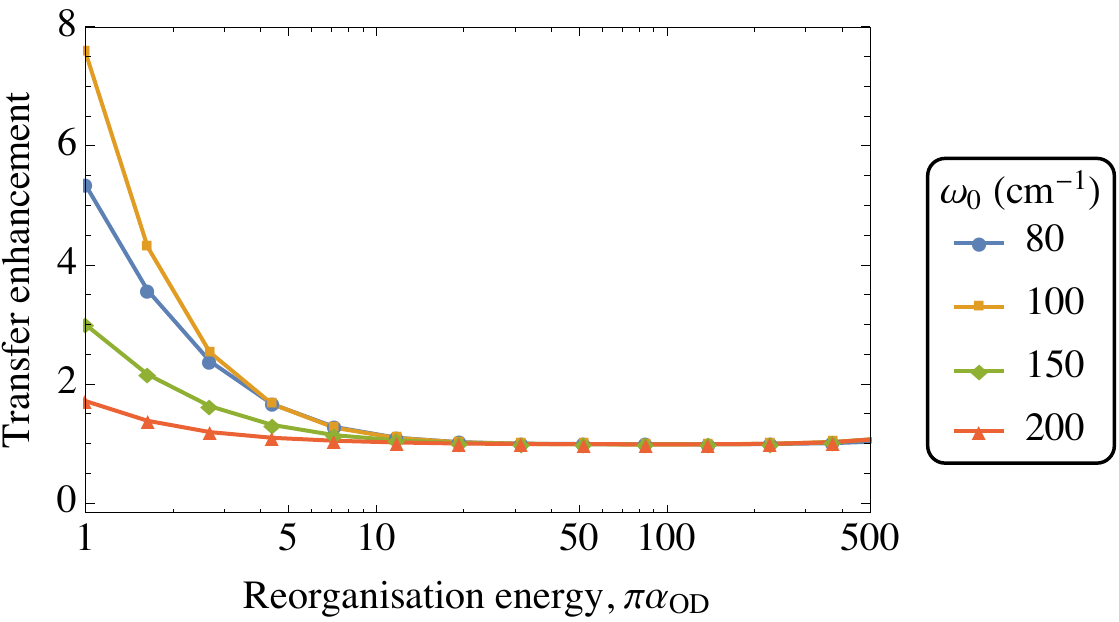}
			\caption[Energy transfer rate for a structured environment.]{Enhancement of the dimer transfer rate due to environmental structure. The reorganisation energy of the underdamped component is kept constant at $\pi\alpha_{\rm UD} = 2$~cm$^{-1}$ and several frequencies $\omega_0$ are considered, while the reorganisation energy of the overdamped component is increased.
			Top: Intersite transfer rate as a function of overdamped reorganisation energy.
			For comparison, we have also included the transfer rate for a single overdamped environment. 
			Bottom: Corresponding transfer enhancement, defined as the ratio of the rate with environmental structure to that of the single overdamped environment. 
			For both plots the dimer and overdamped parameters are the same as Fig.~\ref{fig:rate}. The underdamped environments have $\Gamma=20$~cm$^{-1}$ and $T=300$~K.} 
			\label{fig:st_rate}
\end{figure}

Given that an increase in the rate may be obtained for an underdamped spectrum, it is natural to ask under what circumstances this remains true for a structured environment in which a broad overdamped background is also present (as is likely in any real system). We explore this in Fig.~\ref{fig:st_rate}, where we set the underdamped contribution to a constant reorganisation energy, while the coupling to the overdamped environment is varied. 
Here, we see that the when the overdamped part of the environment is relatively weakly coupled to the dimer, the presence of the underdamped structure can significantly enhance the EET rate even at a temperature of $300$~K. As expected, this is particularly apparent when the underdamped environment is close to the dimer resonance, decreasing as it is tuned away. 
Again, the enhancement is due to the underdamped environment increasing the number of transfer pathways generated by the manifold of vibronic states. 
When 
it is tuned away from resonance, the effective coupling between the dimer and the underdamped RC is reduced, thus decreasing the amount of vibronic states that can be explored by the composite system, and consequently reducing the rate of EET. Note, however, that at large overdamped reorganisation energies the enhancement is suppressed regardless of the underdamped frequency, and the transfer rate follows that of a single overdamped environment. This is true even when the underdamped component can still be clearly discerned within the spectral density (see Fig.~\ref{fig:struct_schem}), and is simply a consequence of the overdamped environment becoming the dominant influence, such that the underdamped vibronic states 
have little effect even for a resonant mode. Thus, the presence of one or more well-resolved modes within the spectral density is not in itself sufficient to imply a vibronic enhancement of the dimer EET rate.

\section{Summary}\label{sec:sec6}
In summary, we have shown that the RCME provides a powerful, informative and intuitive method for describing EET in molecular dimers -- and more generally open quantum systems -- for regimes in which the environmental correlation time is long. It allows access to information on the system, its environment, and their correlations. 
Moreover, it greatly outperforms the closely related semiclassical Zusman equations. This demonstrates that not only is the RC mapping important to capture the correct system behaviour, but also that one must properly account for the correlations dynamically generated between the dimer and its environment through the RC.  
These correlations lead to complex system population dynamics 
comprising of multiple oscillation frequencies, 
which we interpret as feedback from the environmental collective mode. 
They also persist into the steady-state, pushing both the system and its environment away from their respective canonical equilibrium states in the long time limit.

We have applied the RC model to describe the behaviour for overdamped, underdamped, and structured vibrational environments. 
In particular, we find that the presence of structure within the environment is capable of increasing the rate of EET between the dimer sites. 
This enhancement is dependent upon the energy scale of the underdamped vibrations within the environment, reaching its peak when they 
lie close to the dimer excitonic resonance as should be expected. Nevertheless, even for such resonance conditions, there are also 
regions of parameter space in which the structured environment offers no advantage in terms of an increased transfer rate, 
and the dynamics follows that determined by the broad overdamped background. 
It would thus be extremely interesting to apply the RCME to analyse such subtleties in larger molecular systems, with the aid of a suitable truncation scheme to limit the required number of basis states. 

\section{Acknowledgements}

We would like to thank Oliver Woolland, Alex Chin, Brendon Lovett, Zach Blunden-Codd, Dave Newman, Philipp Strasberg and Mauro Cirio for interesting discussions. J.I.-S. is supported by the Danish Research Councils, grant number DFF -- 4181-00416, and A.N. is supported by The University of Manchester. A.G.D. was supported by a Marie Curie International Incoming Fellowship within the 7th European Community Framework Programme. N.L. is partially supported by the FY2015 Incentive Research Project.

\appendix

\section{The Zusman equations}

\label{sec:Zusm}
In this Appendix we shall give further details on the derivation of the Zusman equations. 
Starting from the Caldeira-Leggett master equation in Wigner  space we have~\cite{B614554J}
\begin{align}\label{eq:phase}
\Omega\pi\gamma\frac{\partial}{\partial p}\left(p\hat{W}+\frac{1}{\beta}\frac{\partial\hat{W}}{\partial p}\right)=&\frac{\partial\hat{W}}{\partial t}+\mathcal{H}\hat{W}+i\mathcal{Z}\frac{\partial\hat{W}}{\partial p}\nonumber\\
&+p\frac{\partial\hat{W}}{\partial x}-\Omega^2x\frac{\partial\hat{W}}{\partial p},
\end{align}
where we have defined the superoperators
\begin{align}
\mathcal{H}\hat{W}&=i\left[\left( \frac{\epsilon}{2}+\kappa x\right)\sz+ \frac{\Delta}{2}\sx,\hat{W}\right],\nonumber\\
\mathcal{Z}\frac{\partial\hat{W}}{\partial p}&=\frac{i\kappa}{2}\left\{\sz,\frac{\partial\hat{W}}{\partial p}\right\}.
\end{align}
We aim to simplify this equation of motion by removing the momentum coordinate, in particular by assuming that the momentum of the RC remains in thermal equilibrium throughout the evolution of the system.

To eliminate the momentum coordinate from Eq.~(\ref{eq:phase}) we shall use the procedure outlined by Coffey~\cite{B614554J}, and originally formulated by Brinkman for the case of a Brownian oscillator~\cite{BRINKMAN195629}. We expand the Wigner function in terms of Hermite polynomials
\begin{equation}\label{eq:decomp}
\hat{W}=e^{-\mu^2/4}\sum\limits_{n=0}^{\infty}D_n(\mu)\hat{\phi}_n(x,t),
\end{equation}
where we have rescaled the momentum coordinate such that $\mu=\sqrt{\beta} p$. The function $\hat{\phi}(x,t)$ is a two-by-two matrix describing the electronic dependence of the Wigner function and $D_n(\mu)$ is the set of orthogonal Weber functions, which are given by
$$
D_n(y)=2^{-n/2} e^{-y^2/4}H_n\left(\frac{y}{\sqrt{2}}\right),
$$
where $H_n(z)$ are the Hermite polynomials. The Weber functions satisfy the following relations
\begin{align}\label{eq:web}
D_{n+1}(y)-yD_{n}(y)+n D_{n-1}(y)&=0,\\
\partial_y D_n(y)+\frac{y}{2}D_{n}(y)-n D_{n-1}(y)&=0,\\
\partial_y D_n(y)-\frac{y}{2}D_{n}(y)+ D_{n+1}(y)&=0,\\
\partial^2_y D_n(y)+\left(n+\frac{1}{2}-\frac{y^2}{4}\right)D_n(y)&=0,\\
\int\limits_{-\infty}^{\infty}D_n(y)D_m(y)dy&=n!\sqrt{2\pi}\delta_{m,n}.
\end{align}

Substituting these expressions into the phase-space master equation given in Eq.~(\ref{eq:phase}), and integrating over our scaled momentum coordinate $\mu$, we obtain the Brinkman hierarchy
\begin{align}\label{eq:brink}
\frac{\partial\hat{\phi}_m}{\partial t}+\mathcal{H}\hat{\phi}_m+\frac{1}{\sqrt{\beta}}\left(\frac{\partial\hat{\phi}_{m-1}}{\partial x}+(m+1)\frac{\partial\hat{\phi}_{m+1}}{\partial x}\right)&\nonumber\\
+\sqrt{\beta}\Omega^2 x\hat{\phi}_{m-1}+\pi\gamma\Omega m \hat{\phi}_{m}-\sqrt{\beta}\mathcal{Z}\hat{\phi}_{m-1}&=0.
\end{align}
We also define the differential operators
\begin{equation}
\begin{split}
\mathcal{J}&=-\frac{\sqrt{\beta}}{\eta}\left(\frac{1}{\beta}\frac{\partial}{\partial x}+\Omega^2x-i\mathcal{Z}\right),\\
\mathcal{J}_D&=-\frac{1}{\eta\sqrt{\beta}}\frac{\partial}{\partial x},
\end{split}
\end{equation}
which allows us to write
\begin{equation}\nonumber
\frac{1}{\eta}\left(\dot{\hat{\phi}}_m+\mathcal{H}\hat{\phi}_m\right)+m\hat\phi_m=\mathcal{J}\hat\phi_{m-1}+(m+1)\mathcal{J}_D\hat\phi_{m+1},
\end{equation}
where $\eta=\pi\gamma\Omega$ quantifies the friction acting on the mode. We now move to Laplace space with respect to the time coordinate, using the transformation
\begin{equation}
\tilde\varphi_n=\tilde\varphi_n(x,s)=\int\limits_{-\infty}^{\infty} \hat\phi_n(x,t)e^{-st}dt,\\
\end{equation}
which leads to the relation
\begin{equation}
\frac{\partial\hat\phi_n}{\partial t}\mathrel{\mathop{\Longrightarrow}^{\mathrm{LT}}} s\tilde\varphi_n-\hat\phi_n(x,0).
\end{equation} 
For a mode initially in a thermal state, the initial conditions of the system are entirely determined by $\hat\phi_0(x,0)$, such that $\hat\phi_n(x,0)=0$ for $n>0$, leading to the following hierarchy of equations in Laplace space:
\begin{equation}\nonumber
\begin{split}
&\frac{1}{\eta}\left(s\tilde\varphi_0+\hat\phi_0(x,0)+\mathcal{H}\tilde\varphi_0\right)=\mathcal{J}_D\tilde\varphi_1,\\
&\frac{1}{\eta}\left(s\tilde\varphi_1+\mathcal{H}\tilde\varphi_1\right)+\tilde\varphi_1=\mathcal{J}\tilde\varphi_0+ 2\mathcal{J}_D\tilde\varphi_2,\\
&\frac{1}{\eta}\left(s\tilde\varphi_2+\mathcal{H}\tilde\varphi_2\right)+2\tilde\varphi_2=\mathcal{J}\tilde\varphi_1+ 4\mathcal{J}_D\tilde\varphi_3,\\
&\hspace{3cm}{\vdots}
\end{split}
\end{equation}
We can close these equations by assuming $\tilde\varphi_3=0$ (which is consistent with keeping terms to leading order in $\eta^{-2}$) and hence solve for $\tilde\varphi_0$. Inverting the equation for $\tilde\varphi_2$,
\begin{equation}
\begin{split}
\tilde\varphi_2=&\frac{\mathcal{J}\tilde\varphi_1}{\frac{1}{\eta}(s+\mathcal{H})+2},
\end{split}
\end{equation}
and substituting this into the equation for $\tilde\varphi_1$ gives
\begin{equation}
\tilde\varphi_1=\frac{\mathcal{J}\tilde\varphi_0}{\frac{1}{\eta}(s+\mathcal{H})+1+\frac{2\mathcal{J}_D\mathcal{J}}{\frac{1}{\eta}(s+\mathcal{H})+2}},
\end{equation}
which leads to the equation
\begin{equation*}
\frac{1}{\eta}\left(s\tilde\varphi_0+\hat\phi_0(x,0)+\mathcal{H}\tilde\varphi_0\right)=\frac{\mathcal{J}_D\mathcal{J}\tilde\varphi_0}{1+\frac{1}{\eta}(s+\mathcal{H})+\frac{2\mathcal{J}_D\mathcal{J}}{\frac{1}{\eta}(s+\mathcal{H})+2}}.
\end{equation*}
In the very large damping limit, the friction coefficient $\eta$ is much larger than any other scale. Hence, we can keep terms only to leading order in the inverse friction $\eta^{-1}$, giving
\begin{equation}
s\tilde\varphi_0+\hat\phi_0(x,0)+\mathcal{H}\tilde\varphi_0=\eta \mathcal{J}_D \mathcal{J}\tilde\varphi_0.
\end{equation}
Inverting the Laplace transform we therefore have
\begin{equation}~\label{eq:PhiZus}
\frac{\partial\hat{\phi}_0}{\partial t}=-i\left[\left( \frac{\epsilon}{2}+\kappa x\right)\sz+ \frac{\Delta}{2}\sx,\hat\phi_0\right]+\eta \mathcal{J}_D\mathcal{J}\hat\phi_0.
\end{equation}
Finally, we can decompose $\phi_{0}$ in terms of the dimer states, such that $\hat\phi_0=\sum_{i,j}^2 \mu_{ij}(x,t)\outt{i}{j}$, where $\mu_{ij}(x,t)$ describes an element of the dimer density matrix and is dependent on the RC position.  
Substituting into Eq.~(\ref{eq:PhiZus}) we obtain the Zusman equations

\begin{widetext}
\begin{align}
\frac{\partial\mu_{11}(t,x)}{\partial t}&=\frac{1}{2\pi\gamma\Omega}\frac{\partial}{\partial x}\left((\Omega^2 x+\kappa)\mu_{11}(t,x)+\frac{1}{\beta}\frac{\partial\mu_{11}(t,x)}{\partial x}\right)+i \frac{\Delta}{2}(\mu_{12}(t,x)-\mu_{21}(t,x)),\\
\frac{\partial\mu_{22}(t,x)}{\partial t}&=\frac{1}{2\pi\gamma\Omega}\frac{\partial}{\partial x}\left((\Omega^2 x+\kappa)\mu_{22}(t,x)+\frac{1}{\beta}\frac{\partial\mu_{22}(t,x)}{\partial x}\right)-i \frac{\Delta}{2}(\mu_{12}(t,x)-\mu_{21}(t,x)),\\
\frac{\partial\mu_{12}(t,x)}{\partial t}&=\frac{1}{2\pi\gamma\Omega}\frac{\partial}{\partial x}\left(\frac{1}{\beta}\frac{\partial\mu_{12}(t,x)}{\partial x}+\Omega^2 x\mu_{12}(t,x)\right)
+i(\epsilon +2\kappa x)\mu_{12}(t,x)+i \frac{\Delta}{2}(\mu_{11}(t,x)-\mu_{22}(t,x)),\\
\frac{\partial\mu_{21}(t,x)}{\partial t}&=\frac{1}{2\pi\gamma\Omega}\frac{\partial}{\partial x}\left(\frac{1}{\beta}\frac{\partial\mu_{21}(t,x)}{\partial x}+\Omega^2 x\mu_{21}(t,x)\right)+i(\epsilon +2\kappa x)\mu_{21}(t,x)-i \frac{\Delta}{2}(\mu_{11}(t,x)-\mu_{22}(t,x)).
\end{align}
\end{widetext}
\newpage
To solve the Zusman equations we need to specify initial conditions. If we assume that the system starts in the excited state and the mode in a thermal state $\rho_{\rm th}=\mathcal{Z}_0^{-1}\exp\{-\beta\Omega \ad a\}$, then the only non-zero variable will be $\mu_{11}(x,0)$. Hence, the thermal state in Wigner space may be written as
\begin{equation}
W_{\rm th}=\frac{2 \tanh\left(\frac{\beta\Omega}{2}\right)}{\pi}e^{- \tanh\left(\frac{\beta\Omega}{2}\right)\left(\Omega x^2+\frac{1}{\Omega}p^2\right)}.
\end{equation}
We then integrate over the momentum coordinate to attain the initial condition
\begin{align}\
\mu_{11}(x,0)&=\int\limits_{-\infty}^{\infty}dp W_{th}(0,x,p),\nonumber\\
&=2 \sqrt{\frac{ \tanh\left(\frac{\beta\Omega}{2}\right)}{\pi}}e^{- \Omega\tanh\left(\frac{\beta\Omega}{2}\right) x^2},
\end{align}
while $\mu_{12}(x,0)=\mu_{21}(x,0)=\mu_{22}(x,0)=0$ for all $x$.


\providecommand{\noopsort}[1]{}\providecommand{\singleletter}[1]{#1}%

\end{document}